\documentclass[11pt,a4paper]{article}
\usepackage{jheppub}

\newcommand*{\no}{\noindent}
\newcommand*{\bea}{\begin{eqnarray}}
\newcommand*{\eea}{\end{eqnarray}}
\newcommand*{\be}{\begin{equation}}
\newcommand*{\ee}{\end{equation}}
\newcommand*{\pd}{\partial}
\newcommand*{\pdm}{\pd_{\mu}}
\newcommand*{\pdn}{\pd_{\nu}}

\newcommand*{\pref}[1]{(\ref{#1})}

\newcommand*{\mn}{{\mu\nu}}

\newcommand*{\nn}{\nonumber}

\newcommand*{\tr}{\mathrm{tr}}

\newcommand*{\indexsep}{,}

\title{On the gauge-algebra dependence of Landau-gauge Yang-Mills propagators}

\author{Axel Maas\footnote{Present address: Institute for Theoretical Physics, Friedrich-Schiller-University Jena, Max-Wien-Platz 1, D-07743 Jena, Germany}}

\affiliation{Department of Theoretical Physics, Institute of Physics, Karl-Franzens University Graz,\\ Universit\"atsplatz 5, A-8010 Graz, Austria E-Mail}

\emailAdd{axelmaas@web.de}

\abstract{Yang-Mills theory can be formulated for any semi-simple Lie-algebra, and thus any semi-simple Lie-group. In principle, the dynamics could be different for each one. However, functional studies predict that the propagators in Landau gauge depend only quantitatively on the gauge algebra. In particular, genuine non-perturbative effects should be present even in the large $N$-limit for su($N$) gauge algebras.

Lattice gauge theory is used to investigate this in detail. The propagators are determined for the gauge groups SU(2), SU(3), SU(4), SU(5), SU(6) and G$_2$, in two and three dimensions. In accordance with the prediction no qualitative dependence on the gauge group is found. In particular, no diminishing of non-perturbative contributions is found for $N$ becoming large in the SU($N$) case. Quantitative effects are found, and analyzed in detail.}


\begin{document}

\maketitle


\section{Introduction}

Yang-Mills theories are gauge theories of (semi-)simple Lie-algebra valued gauge fields $A_\mu^a$, governed by the Lagrangian \cite{Bohm:2001yx}
\bea
{\cal L}&=&-\frac{1}{4}F^{\mn a} F_\mn^a\nn\\
F^a_\mn&=&\pdm A_\nu^a-\pdn A_\mu^a-gf^{abc}A_\mu^bA_\nu^c\nn,
\eea
\no where the $f^{abc}$ are the structure constants of the chosen gauge-algebra. The coupling constant $g$ determines, in four dimensions via dimensional transmutation, in other dimensions directly, the scale of the theory. The gauge field themselves belong to the adjoint representation of the gauge algebra.

It is always possible to formulate Yang-Mills theories in terms of group-valued variables. However, at face value all group representations for a given gauge algebra would provide the same dynamics on the level of the gluon fields. E.\ g., for the case of the su($N$) algebra, a representation in terms of the groups SU($N$) and SU($N$)/Z$_N$ must be equivalent. This is trivial so, as both representations differ only by group-elements, the center, which leave the gauge fields themselves invariant: Center transformations act as identity transformations on the gluons. However, though the dynamics is necessarily the same, it is of course possible to construct group-valued quantities, which will not be the same in both cases, e.\ g., Polyakov loops\footnote{In case of semi-simple Lie groups, this can become even more complicated, since the algebra is just a product-algebra, while this is not necessarily the case for the corresponding groups \cite{O'Raifeartaigh:1986vq,vonSmekal:2010la}.}. Such constructions become even even more important when the theories are formulated on a non-trivial space-time manifold, like a torus in lattice gauge theory or at finite temperature in equilibrium. Anomalies due to matter fields can restrict the possible group representations further. In case of the standard model actually the only permitted group representation of the gauge algebra su(3)$\times$su(2)$\times$u(1) is S(U(3)$\times$U(2)). This yields the unbroken gauge group SU(3)/Z$_3\times$U(1) of the electrostrong interactions and the broken SU(2)/Z$_2$ of the weak isospin \cite{O'Raifeartaigh:1986vq}.

However, the dynamics could be very different for different gauge algebras. In perturbation theory, this is evidently not the case: The different algebras only manifest themselves in different coefficients of the expansion, without altering qualitatively the behavior \cite{Bohm:2001yx}. In particular, for any Lie-algebra Yang-Mills theories are (in four dimensions) renormalizable with the same number of independent renormalization constants, and are all asymptotically free. This property is not remaining beyond perturbation theory. E.\ g., the order of the finite-temperature phase transition depends on the chosen gauge algebra, as already the comparison of su(2) and su(3) in terms of the groups of SU(2) and SU(3) shows \cite{Lucini:2005vg}: In one case the transition is second order, in the other first order. Also the bound state spectrum is different due to algebraic reasons. E.\ g., su(2) is not having negative charge-conjugation-parity glueball bound states due to the existence of real representations, while su(3) possesses this type of bound states \cite{Lucini:2001ej}.

As soon as matter fields in any representation are coupled to the theory the behavior can even change totally. Even asymptotic freedom is no longer guaranteed if a sufficient number of additional matter fields in appropriate representations are added to the theory. This much more complicated problem will not be treated here, and only the pure Yang-Mills case will be studied.

The main focus of interest here is whether the gluons show different behavior for differing gauge algebras. This will be investigated using lattice gauge theory, by implementing particular gauge groups, and utilizing the gauge-dependent gluon correlation functions. This complements investigations of gauge-independent quantities, like the glueball spectrum \cite{Meyer:2002mk}, string tensions \cite{Teper:2009uf}, or the thermodynamic properties \cite{Liddle:2008kk,Holland:2003kg,Pepe:2006er,Greensite:2006sm}. The advantage of investigating the gauge-dependent correlation functions is that they can be assembled to obtain gauge-invariant quantities, see e.\ g.\ \cite{Alkofer:2000wg,Fischer:2006ub,Roberts:2000aa}, in analogy to perturbation theory. This is an approach which can be applied even were pure lattice calculations are yet restricted, like the chiral limit \cite{Alkofer:2000wg,Fischer:2006ub}, cold and dense matter \cite{Roberts:2000aa,Nickel:2008ef,Marhauser:2006hy}, or cases with disparate characteristic scales \cite{Alkofer:2000wg}. Analyzing the group-dependence of the correlation functions is then an important input to improve the systematics of the approach and of the assumptions made.

Since the properties of gluons are gauge-dependent this has to be done in a fixed gauge. This will be chosen here to be the Landau gauge. In the non-perturbative domain this is not a sufficiently precise definition of the gauge due to the presence of Gribov-Singer copies \cite{Gribov:1977wm,Singer:1978dk}. For the present purpose the minimal \cite{Cucchieri:1995pn} or average-B \cite{Maas:2009se} Landau gauge as the sub-type of non-perturbative Landau gauges will be chosen. However, since the calculation of large gauge-algebras becomes exceedingly expensive when it comes to gauge-fixing, only rather small volumes will be studied. For these volumes, a significant effect of Gribov copies for the gluon is not expected, and only a moderate effect for the ghost \cite{Maas:2009se,Maas:2008ri,Bornyakov:2008yx}.

Furthermore, due to the numerical costs for the calculations, here only the case for two and three dimensions will be investigated, and only the propagators will be determined. In case of the gauge groups SU(2) and SU(3) also comparisons in four dimensions have been performed \cite{Bogolubsky:2009dc,Cucchieri:2008fc,Cucchieri:2007rg,Sternbeck:2007ug,Oliveira:2008uf}. These showed no essential differences in both cases\footnote{See, however, \cite{Oliveira:2008uf} for a differing conclusion in four dimensions.}. As this agrees with the findings here, it is conceivable that also the results for other gauge groups can be translated from lower dimensions to four dimensions, though this conjecture requires confirmation. The set of groups covered here is SU(2), SU(3), SU(4), SU(5), SU(6), and G$_2$.

The ladder of SU($N$) groups permits to investigate also the possibility to which extent at fixed 't Hooft coupling $\lambda=g^2N$ the results show a simple dependence on $N$ with increasing $N$. Indeed, many results on gauge-invariant quantities seem to indicate such a behavior, see e.\ g.\ \cite{Lucini:2005vg,Lucini:2001ej,Meyer:2002mk,Bringoltz:2005rr,Lucini:2004yh,Witten:1979kh}. All of these quantities have a genuine non-perturbative origin, and require confinement as well as strong interactions for the bound states to exist \cite{Witten:1979kh}, and therefore indicate that the large-$N$ limit is non-perturbative. This is to be expected, already based on Haag's theorem \cite{Haag:1992hx}, irrespective of whether the possible range of possible $g$ values, and therefore of $\lambda$, is $N$-dependent or not. Here, evidence will be provided that the gauge-dependent correlation functions are essentially independent of $N$ at fixed $\lambda$, confirming that the large-$N$ limit is highly non-perturbative, but exhibiting a simple scaling with $N$, indeed in this case a trivial one.

The motivation to study G$_2$ is different. Since all realizations of the g$_2$ algebra have a trivial center, many arguments based on degrees of freedom sensitive to the center of the gauge group do not apply \cite{Greensite:2006sm,Holland:2003jy}. Nonetheless, the dynamics of G$_2$ gauge theories appears to be quite similar to theories with a center \cite{Pepe:2006er,Greensite:2006sm,Maas:2007af,Cossu:2007dk,Danzer:2008bk,Wellegehausen:2009rq,Wellegehausen:2010ai}. This observation will also be confirmed here. To understand this result and its implications is therefore helpful to identify the relevant effective degrees of freedom in gauge theories. In a sense, since g$_2$ has a rather different group-theoretical structure, it tests the extreme case of choosing a different gauge algebra. The investigation of G$_2$ Landau-gauge propagators here extends previous studies \cite{Maas:2007af}.

The insensitivity on the gauge group and the non-triviality of the $N\to\infty$ limit are actually not a surprise, and have been pointed out in functional calculations \cite{vonSmekal:1997is,vonSmekal:1997vx,Maas:2005ym}. These arguments will be rehearsed in subsection \ref{sdse}. In the corresponding section \ref{ssce} also other aspects of the continuum theory will be discussed. However, calculations using functional methods involve truncations, and therefore the results have to be tested. This is the main aim here, to test these predictions using the complementary approach of lattice gauge theory. The setup for these lattice calculations is given in section \ref{slat} and \ref{sprop}. The results, demonstrating indeed a qualitative gauge-group independence, are given in section \ref{sresults}. A short summary will be provided in section \ref{ssum}.

\section{The asymptotic gauge-group dependence}\label{ssce}

\subsection{Setup}

Continuum (Euclidean) Yang-Mills theory in (perturbative) Landau gauge is formulated as the Lagrangian
\bea
{\cal L}&=&\frac{1}{4}F_\mn^a F_\mn^a+\bar c^a\pdm D^{ab}_\mu c^b\nn\\
F^a_\mn&=&\pdm A_\nu^a-\pdn A_\mu^a-gf^{abc}A_\mu^bA_\nu^c\nn\\
D_\mu^{ab}&=&\delta^{ab}\pdm+gf^{abc}A_\mu^c.\label{lag}
\eea
\no Herein are $c^a$ ($\bar c^a$) the (anti-)ghost field, and the index $a$ on both fields counts the dimensionality of the adjoint representation of the gauge group.

To complete the gauge non-perturbatively, it is necessary to chose between non-perturbative realizations \cite{Gribov:1977wm,Singer:1978dk}. Arguments have been provided that this can be done by imposing conditions on certain correlation functions \cite{Maas:2009se,Maas:2008ri,Fischer:2008uz}, e.\ g.\ on the ghost propagator \cite{Maas:2009se,Fischer:2008uz}. However, this issue is not yet fully settled, see \cite{Maas:2010wb} for a status report.

Here, for the sake of simplicity, it will be assumed that the construction of \cite{Maas:2009se,Fischer:2008uz} is indeed possible. Then, a gauge choice can be made such that the ghost dressing function $p^2 D_G(p)$, where $D_G$ is the ghost propagator, is infrared singular. This permits a simpler calculation when using functional equations below. A similar line of argumentation can be expected to hold also in photon-ghost gauges, where $p^2 D_G(p)$ is chosen to be finite, though this has yet to be done explicitly. For the lattice calculations presented in section \ref{slat}, it can be expected that the choice of non-perturbative Landau gauge is not relevant for the gauge-group dependence, in particular due to the limited volumes accessible. Therefore, the numerically much cheaper minimal or average-B Landau gauge is used \cite{Cucchieri:1995pn,Maas:2009se}, which belongs to the class of photon-ghost gauges.

One possibility to obtain from the Lagrangian \pref{lag} the propagators are the Dyson-Schwinger equations (DSEs) \cite{Alkofer:2000wg}. Keeping all color indices and the structure constants explicit, the Dyson-Schwinger equations are form-invariant for all gauge groups. In particular, only the structure constants $f^{abc}$ do appear. Additional structure constants, like the symmetric one $d^{abc}$, being zero for most gauge groups \cite{Cvitanovic:2008}, do not appear explicitly in the DSEs. Of course, the full Green's functions can develop such contributions.

The explicit derivation of the DSEs is discussed in great detail elsewhere \cite{Alkofer:2000wg}. Only two particular limits will be of interest here. One is the far infrared limit, and the other the far ultraviolet limit. In the latter case, the equations generate (resummed) perturbation theory.

The following discussion is presented for completeness, and to emphasize the role of the gauge group. It follows previous presentations of the subject \cite{vonSmekal:1997vx,Maas:2005ym}.

\subsection{The far ultraviolet}\label{suvpert}

There is a qualitative difference between four-dimensional and lower-dimensional systems: Irrespective of the gauge group neither in two, nor in three dimensions any physical renormalization occurs\footnote{Note that regularization is still necessary. However, a standard BPHZ-like prescription would permit to perform this implicitly \cite{Collins:1984xc}.}.

In four dimensions the propagators behave at large momenta to leading order like 
\be
p^2 D(p)\stackrel{p\gg \Lambda_{\mathrm{YM}}}{\to}(1+\omega\ln p)^\delta,\no
\ee
\no where $\omega$ and $\delta$ depend on the gauge group, but have the same sign for all gauge groups, and $\Lambda_\mathrm{YM}$ is the scale generated by dimensional transmutation. In two and three dimensions, however, the propagators depend only polynomial on the momentum. In particular, no resummation occurs to leading order in $g$, and only starting from next-to-leading order such effects appear \cite{Maas:2004se}. Hence, the propagator take at large momenta the form
\be
p^2 D(p)\stackrel{p\gg g^2}{\to}1+c\frac{g^2 C_A}{p}\label{uv1}
\ee
\no in three dimensions and
\be
p^2 D(p)\stackrel{p\gg g}{\to}1+c\frac{g^2 C_A}{p^2}\label{uv2}
\ee
\no in two dimensions, just on dimensional grounds. $C_A$ is the adjoint Casimir of the gauge group, defined by
\be
C_A \delta^{ab}=f^{acd}f^{bcd}\nn.
\ee 
\no The constants of proportionality $c$ depend only on the underlying space-time manifold. In three dimensions they take the value $11/64$ for the gluon propagator and $1/16$ for the ghost propagator. In two dimensions infrared divergencies appear, which make a somewhat ad-hoc regularization necessary\footnote{Strictly speaking, this problems occur also in three dimensions, though only at higher order \cite{Jackiw:1980kv}. I am grateful to David Dudal for pointing this out.}. This already implies that non-perturbative contributions will appear which will provide such a regularization. It turns out that this is due to an infrared suppression, compared to tree-level, of the gluon propagator \cite{Lerche:2002ep,Zwanziger:2001kw,Maas:2007uv,Pawlowski:2003hq}. This makes the results convergent.

Hence, already in perturbation theory the gluon (and ghost) propagator differ for the different algebras. E.g., the leading order perturbative coefficient changes by the ratio of the adjoint Casimir operator from gauge algebra to gauge algebra in two and three dimensions. However, the momentum dependence to first order stays the same. Therefore, perturbatively, the gauge algebra dependence is only quantitative. Note that the perturbative expansion does not depend on the chosen group representation of the algebra.

\subsection{The far infrared}\label{sdse}

For the sake of simplicity the argument why the gauge-algebra dependence in the non-perturbative domain should also be only a quantitative rather than a qualitative effect, the following discussion will be done in a scaling-type gauge \cite{Maas:2009se,Fischer:2008uz}. This section essentially follows the line of arguments of \cite{Maas:2005ym}. The situation in a photon-ghost gauge \cite{Fischer:2008uz,Binosi:2009qm,Aguilar:2006gr,Dudal:2008sp,Boucaud:2008ji}, like the one used for the lattice calculations, will be discussed below.

Here, only the leading infrared contributions will be retained, a truncation discussed extensively elsewhere \cite{Fischer:2008uz,Lerche:2002ep,Zwanziger:2001kw,Pawlowski:2003hq,Alkofer:2004it,Fischer:2006vf,Fischer:2009tn,Huber:2007kc}. This truncation requires to keep only terms up to one-loop, which include at least one ghost line. Therefore, also the gluon tree-level term can be dropped self-consistently. The equations then read
\bea
D^{ab-1}_G(p)&=&-\widetilde Z_3\delta^{ab}p^2\label{gheq}\\
&&+gf^{abc}\int\frac{d^dq}{(2\pi)^d}ip_\mu(-q,p,q-p)D^{ef}_{\mu\nu}(p-q)D^{dg}_G(q)\Gamma^{c\bar cA\indexsep bgf}_\nu(-p,q,p-q)\nn\\
D^{ab-1}_{\mu\nu}(p)&=&-gf^{abc}\int\frac{d^dq}{(2\pi)^d}ip_\mu D_G^{cf}(q) D_G^{de}(p+q) \Gamma_\nu^{c\bar c A\indexsep feb}(-q,p+q,-p)\nn,
\eea
\no where $d$ is the dimensionality, $D_G$ is the ghost propagator, $D_\mn$ is the gluon propagator, $\widetilde Z_3$ is the possibly finite ghost wave function renormalization, and $\Gamma_\nu^{c\bar c A\indexsep abc}$ is the full ghost-gluon vertex. The latter is undetermined at this level of the truncation, and will be set to its bare counterpart. For SU(2) and SU(3) this appears to be an even quantitatively good approximation \cite{Maas:2007uv,Cucchieri:2008qm,Ilgenfritz:2006he}, and is at least self-consistent for arbitrary gauge groups \cite{Fischer:2006vf,Fischer:2009tn,Huber:2007kc,Schleifenbaum:2004id}. If an additional color-tensor proportional to the symmetric color tensor $d^{abc}$ would appear, this would get lost due to the contraction with the antisymmetric tree-level vertex. Hence, at least the leading infrared part will have for neither propagator color off-diagonal elements. Only at two-loop order, and thus infrared and ultraviolet sub-leading, this could happen. The results on the lattice presented in section \ref{sresults} and most accurate in this intermediate energy domain show that this is not the case.

Taking the propagators to be of the form
\bea
D^{ab}_G(p)&=&-\delta^{ab}A_Gp^{-2-2\kappa}\label{ir1}\\
D^{ab}_\mn(p)&=&-\delta^{ab}\left(\delta_\mn-\frac{p_\mu p_\nu}{p^2}\right)A_Zp^{-2-2t}\label{ir2}
\eea
\no it is possible to absorb the remaining tree-level term in \pref{gheq} by implementing the boundary condition \cite{Fischer:2008uz} of an infrared divergent ghost dressing function, i.\ e., setting $1/B=0$ in the language of \cite{Maas:2009se}. The integrals can then be performed analytically to yield
\bea
p^{2\kappa}&=&g^2 C_A A_G^2 A_Z I_G(\kappa,t,d) p^{-(4-d)-2\kappa-2t}\nn\\
p^{2t}&=&g^2 C_A A_G^2 A_Z I_Z(\kappa,d) p^{-(4-d)-4\kappa}\nn.
\eea
\no The expressions $I_G$ and $I_Z$ are functions depending solely on the exponents $\kappa$ and $t$, and the structure of the underlying space-time manifold, symbolized by the dependence on $d$, and can be found, e.\ g., in \cite{Zwanziger:2001kw}. Counting powers of momentum yields the relation
\be
-2\kappa=t+\frac{4-d}{2}\label{conc}.
\ee
\no The remaining consistency condition
\be
I_Z(\kappa,d)=I_G(\kappa,t(\kappa),d)\nn
\ee
\no implies that the exponents are only depending on the space-time manifold, but not on the gauge algebra: At this level, the fact that the propagators behave like a power-law and the value of the exponent is independent of the gauge algebra. A dependence on the gauge algebra can only be induced if the ghost-gluon vertex would be different from the tree-level version, and at least quantitatively dependent on the gauge algebra. However, as long as a solution to \pref{conc} exists with $\kappa\ge 1/2$, this would only affect the numerical value of the exponent, but not the qualitative behavior of the propagators\footnote{Note that $\kappa=1/2$ corresponds to an infrared finite rather than an infrared vanishing gluon propagator. This is not a qualitative different solution, when the characteristic of the solution is taken to be the relation \pref{conc}, rather than the individual behavior of the propagators \cite{Fischer:2006vf,Fischer:2009tn}.}. This yields the conjecture of the qualitative independence of the infrared behavior of the propagators of the gauge algebra.

The pre-factors are, however, gauge-algebra dependent by virtue of the equation
\be
g^2C_A A_G^2A_Z=\frac{1}{I_Z(\kappa,d)}=\frac{1}{I_G(\kappa,\kappa(t),d)}\label{ircoeff}.
\ee
\no This gives the further prediction of a scaling relation with the adjoint Casimir in the far infrared
\be
p^{2+d}D_G^{aa2}D_{\mu\mu}^{aa}\sim \frac{1}{g^2 C_A}\label{scagg}.
\ee
\no In particular, in the 't Hooft limit \cite{'tHooft:1973jz} of su($N$) Yang-Mills theory the left-hand side of \pref{scagg} must be a constant as a function of the number of colors $N$. Hence, $A_G$ and $A_Z$ must both scale in some way as a function of $g^2C_A$ to ensure \pref{scagg}. However, if the 't Hooft-limit is taken, both $A_G$ and $A_Z$ must either compensate their respective scaling behavior, or must be individually independent of $N$. The latter behavior is indeed observed in the lattice calculations later, notably in two dimensions where a scaling behavior is manifest. This implies that the effective coupling, given by \cite{vonSmekal:1997vx}
\be
\alpha(p^2)=g^2C_Ap^{6}D_G^{aa2}D_{\mu\mu}^{aa}\label{coupling},
\ee
\no will be essentially independent of the gauge algebra, and any dependence can only be introduced by the experimental input $\alpha(\mu^2)$, which cannot be fixed inside the theory.

Note that the propagators are still non-trivial in the $N\to\infty$ limit, and their properties are not obtainable by perturbative calculations in the planar limit. Nonetheless, the non-perturbative planar limit of the Dyson-Schwinger equations is sufficient to obtain this result. The failure of perturbation theory in the large $N$-limit is expected, as argued in the introduction. However, the fact that a non-perturbative planar limit of the DSEs is sufficient for the asymptotic infrared behavior is a non-trivial result. This actually generalizes to the DSEs for any Green's functions \cite{Alkofer:2004it,Fischer:2006vf,Fischer:2009tn,Huber:2007kc}. Note that also the all-order power-counting analysis \cite{Fischer:2006vf,Fischer:2009tn} of DSEs and the exact renormalization group equations still obey the gauge-algebra independence trivially.

Hence, at this level neither $\kappa$ nor $g^2 C_A A_G^2 A_Z$ receive any $1/N$ corrections in the large $N$-limit. Any dependence of these quantities on $N$ therefore indicate $N$-dependent vertex corrections.

When moving to the photon-ghost case, the situation is less simple to access. In this case, at low momenta all diagrams contribute equally \cite{Maas:2004se,Binosi:2009qm,Dudal:2008rm,Dudal:2008sp,Dudal:2008xd} to the effective gluon screening mass and the effective ghost wave-function dressing. In particular, the two-loop diagrams, which contain four-gluon vertices, potentially provide an equally important contribution. By this, further dependencies on further invariant tensors of the gauge algebra may appear. However, this is again not a qualitative effect, since the infrared behavior is not altered, but it can be quantitatively significant. However, the lattice results below suggest that such quantitative effects are comparatively small.

\section{Lattice formulation}\label{slat}

The standard Wilson action \cite{Gattringer:2010zz}
\be
S=\beta\sum\left(1-\frac{1}{N_F}\Re\tr U_\mn\right)\nn,
\ee
\no is valid for any Lie-group, provided the link matrices $U_\mu$, building the standard plaquette $U_\mn$, are given in the fundamental representation of the group used to represent the gauge algebra. $N_F$ is then the dimension of the fundamental representation, and the sum is over all plaquettes. The bare coupling constant $g$ is encoded in the constant $\beta$ as 
\be
\beta=\frac{2N_F}{g^2 a^{4-d}}\quad\Leftrightarrow \quad g=\left(\frac{2N_F}{\beta a^{4-d}}\right)^{\frac{1}{2}}\nn,
\ee
\no where $a$ is the lattice-spacing. In the implementation employed here, the fundamental representation for the SU($N$) groups has been taken from \cite{Georgi:1982jb}, and for G$_2$ the Macfarlane representation \cite{Macfarlane:2002hr} has been used. $N_F$ takes then the values 2 to 6 for SU(2) to SU(6), and 7 for G$_2$, respectively.

A significant problem in comparing the results for various gauge groups are the potentially different scales \cite{Maas:2007af}. This can be solved most easily by expressing all quantities by dimensionless ratios. An alternative is setting the string tension to the same value, here chosen conventionally to be (440 MeV)$^2$. In this case all quantities are expressed effectively in units of the string tension. It is then possible to select also the same physical volume by choosing the same extension in lattice units and $\beta$ such that for all gauge groups the physical volumes in units of the string tension agree. However, for $d\ge 3$, the asymptotic string tension in case of the gauge group G$_2$ vanishes \cite{Holland:2003jy}. Nonetheless, an intermediate string tension, which is also used for setting the scale in case of the gauge group SU($N$)\footnote{The asymptotic string-tension is the one in the $N$-ality regime \cite{Greensite:2003bk}, which is not necessarily the one observed at short distances of a few fermi.}, is non-vanishing \cite{Greensite:2006sm,Liptak:2008gx}, and will be used here for this purpose.

Unfortunately, even a direct measurement of the intermediate distance string tension is difficult and nontrivial in three and higher dimensions, because various corrections to the string tension are present. The values for SU($N$) have been taken from \cite{Teper:1998te}\footnote{Interpolated and extrapolated in $\beta$ and $N$, where necessary.}. For G$_2$ similar results in three dimensions are by now available in \cite{Wellegehausen:2010ai}, and agree reasonably with the estimation procedure described in \cite{Maas:2007af}, which was used here. In two dimensions, the string tension is determined by direct measurements, which reproduces the known results for SU(2) \cite{Dosch:1978jt}, and supports the method of \cite{Maas:2007af} for G$_2$.

\begin{table}
\caption{\label{beta}The $\beta$ values used in the simulations, together with the corresponding values of $a$ and $g$ for a string tension of (440 MeV)$^2$. $N$ is the size of the largest available volume in lattice units. $P$ is the value of the plaquette from this volume. Note that in two dimensions the same value of the string tension implies the same value of the plaquette. Of course, this would require precisely matching $\beta$ values. Here, an agreement to two digits in $a$ was taken to be sufficient.}
\vspace{1mm}
\begin{tabular}{|c|c|c|c|c|c|c|c|c|c|}
\hline
$d$ & Group & $\beta$ & $a$ [fm] & $a^{-1}$ [GeV] & $g$ [GeV$^\frac{4-d}{2}$] & $C_A$ & $g^2 C_A$ [GeV$^{4-d}$] & $N$ & $P$        \cr
\hline
\hline
2   & SU(2) & 10      & 0.18     & 1.1            & 0.70                      & 2     & 0.97                  & 150 & 0.854185(2)  \cr
\hline
2   & SU(2) & 38.7    & 0.089    & 2.2            & 0.72                      & 2     & 1.0                   & 150 & 0.9614970(7) \cr
\hline
2   & SU(3) & 28      & 0.18     & 1.1            & 0.51                      & 3     & 0.78                  & 112  & 0.85865(2)   \cr
\hline
2   & SU(3) & 100     & 0.089    & 2.2            & 0.54                      & 3     & 0.89                  & 112  & 0.96011(2)  \cr
\hline
2   & SU(4) & 52.9    & 0.18     & 1.1            & 0.43                      & 4     & 0.73                  & 104  & 0.85903(2)   \cr
\hline
2   & SU(4) & 182     & 0.089    & 2.2            & 0.46                      & 4     & 0.86                  & 104  & 0.95885(4)  \cr
\hline
2   & SU(5) & 84.7    & 0.18     & 1.1            & 0.38                      & 5     & 0.71                  & 104  & 0.85882(2)   \cr
\hline
2   & SU(5) & 302     & 0.089    & 2.2            & 0.40                      & 5     & 0.81                  & 104  & 0.960298(4)  \cr
\hline
2   & SU(6) & 117     & 0.18     & 1.1            & 0.35                      & 6     & 0.74                  & 44  & 0.85085(3)   \cr
\hline
2   & SU(6) & 442     & 0.089    & 2.2            & 0.36                      & 6     & 0.79                  & 44  & 0.960425(4)  \cr
\hline
2   & G$_2$ & 50      & 0.18     & 1.1            & 0.58                      & 2     & 0.68                  & 56  & 0.86010(4)   \cr
\hline
2   & G$_2$ & 175     & 0.089    & 2.2            & 0.63                      & 2     & 0.78                  & 56  & 0.959993(5)  \cr
\hline
\hline
3   & SU(2) & 4.24    & 0.17     & 1.2            & 1.0                       & 2     & 2.1                   & 64  & 0.744563(1)  \cr
\hline
3   & SU(2) & 7.09    & 0.094    & 2.1            & 1.1                       & 2     & 2.3                   & 64  & 0.8531134(7) \cr
\hline
3   & SU(3) & 10.7    & 0.17     & 1.2            & 0.81                      & 3     & 1.9                   & 56  & 0.72480(2)   \cr
\hline
3   & SU(3) & 18      & 0.094    & 2.1            & 0.84                      & 3     & 2.1                   & 56  & 0.844316(9)  \cr
\hline
3   & SU(4) & 19.85   & 0.17     & 1.2            & 0.68                      & 4     & 1.9                   & 40  & 0.720095(7)  \cr
\hline
3   & SU(4) & 34.2    & 0.094    & 2.1            & 0.70                      & 4     & 2.0                   & 40  & 0.846161(4)  \cr
\hline
3   & SU(5) & 31.45   & 0.17     & 1.2            & 0.60                      & 5     & 1.8                   & 32  & 0.716233(8)   \cr
\hline
3   & SU(5) & 54.3    & 0.094    & 2.1            & 0.62                      & 5     & 1.9                   & 32  & 0.844750(5)  \cr
\hline
3   & SU(6) & 45.7    & 0.17     & 1.2            & 0.55                      & 6     & 1.8                   & 20   & 0.71469(3)   \cr
\hline
3   & SU(6) & 79.3    & 0.094    & 2.1            & 0.56                      & 6     & 1.9                   & 20  & 0.84491(1)   \cr
\hline
3   & G$_2$ & 18.8    & 0.17     & 1.2            & 0.93                      & 2     & 1.7                   & 24  & 0.72395(3)   \cr
\hline
3   & G$_2$ & 32.9    & 0.094    & 2.1            & 0.94                      & 2     & 1.8                   & 24  & 0.850734(8)  \cr
\hline
\end{tabular}
\end{table}

Still, a quantitative comparison between the different gauge groups makes only sense for inherently dimensionless quantities, like the infrared exponents, as the value of the string tension is arbitrarily set to the same value. Hence, only qualitative statements for dimensionful quantities are sensible. Note that this might even affect the determination of the volumes. The selected values for $\beta$, and the corresponding values of $a$ and $g$, are given in table \ref{beta}. Since the coupling is the only unique dimensionful quantity in these theories, it would be equally well valid to set it to some fixed value, say 1 GeV, and determine by this condition $a$. As the values in table \ref{beta} indicate, this would modify the scales by up to 50\%.

\begin{table}
\caption{\label{conf} The number of configurations for the various systems investigated. The number of configurations is given for the various gauge groups. Swe.\ is the number of sweeps between two consecutive measurements, and Ther.\ is the number of initial thermalization sweeps. Note that always multiple independent runs have been performed to obtain the final statistics. The latter was selected to obtain the ghost exponent with a statistical accuracy of 10\% at the 1$\sigma$-level, computational resources permitting. High and low in the $\beta$ column corresponds to the higher and lower $\beta$ value for the different gauge groups. An empty field indicates that no measurements have been performed for this gauge group at this volume and $\beta$ value. $V$ is the physical volume and $N$ the lattice extension. The data for SU(2) are taken from \cite{Maas:2008ri}.}
\vspace{1mm}
\begin{tabular}{|c|c|c|c|c|c|c|c|c|c|c|c|}
\hline
$d$ & $\beta$ & $N$ & $V^{\frac{1}{d}}$ [fm] & Ther.\ & Swe.\ & SU(2) & SU(3) & SU(4) & SU(5) & SU(6) & G$_2$ \cr
\hline
2   & High    & 4   & 0.36                   & 140    & 14    & 1045  & 935   & 777   & 620   & 465   & 620   \cr
\hline
2   & Low     & 4   & 0.72                   & 140    & 14    & 1045  & 935   & 876   & 620   & 459   & 620   \cr
\hline
2   & High    & 10  & 0.89                   & 200    & 20    & 1589  & 1087  & 840   & 567   & 483   & 543   \cr
\hline
2   & High    & 16  & 1.4                    & 260    & 26    & 1017  & 895   & 735   & 641   & 451   & 597   \cr
\hline
2   & Low     & 10  & 1.8                    & 200    & 20    & 2365  & 1169  & 848   & 567   & 519   & 584   \cr
\hline
2   & High    & 20  & 1.8                    & 300    & 30    & 1094  & 895   & 744   & 642   & 421   & 562   \cr
\hline
2   & High    & 26  & 2.3                    & 360    & 36    & 1015  & 903   & 807   & 629   & 442   & 519   \cr
\hline
2   & Low     & 16  & 2.9                    & 260    & 26    & 1017  & 898   & 734   & 635   & 403   & 610   \cr
\hline
2   & High    & 34  & 3.0                    & 440    & 44    & 1038  & 879   & 712   & 620   & 496   & 566   \cr
\hline
2   & Low     & 20  & 3.6                    & 300    & 30    & 1041  & 896   & 788   & 626   & 433   & 570   \cr
\hline
2   & High    & 44  & 3.9                    & 540    & 54    & 1060  & 931   & 769   & 581   & 147   & 265   \cr
\hline
2   & Low     & 26  & 4.7                    & 360    & 36    & 1009  & 903   & 774   & 581   &       & 554   \cr
\hline
2   & High    & 56  & 5.0                    & 660    & 66    & 1054  & 912   & 718   & 359   &       & 68    \cr
\hline
2   & Low     & 34  & 6.1                    & 440    & 44    & 1038  & 1756  & 795   & 575   & 380   & 552   \cr
\hline
2   & High    & 68  & 6.1                    & 780    & 78    & 1011  & 918   & 516   & 25    &       &       \cr
\hline
2   & High    & 80  & 7.1                    & 900    & 90    & 1113  & 358   & 94    & 51    &       &       \cr
\hline
2   & Low     & 44  & 7.9                    & 540    & 54    & 1059  & 9954  & 1476  &  751  & 59    & 318   \cr
\hline
2   & High    & 104 & 9.3                    & 1140   & 114   & 1045  & 30    & 41    &  25   &       &       \cr
\hline
2   & Low     & 56  & 10                     & 660    & 66    & 6564  & 4611  &  853  & 491   &       & 19    \cr
\hline
2   & High    & 112 & 10                     & 1220   & 122   & 3524  & 10    &       &       &       &       \cr
\hline
2   & Low     & 68  & 12                     & 780    & 78    & 1011  & 1694  & 342   & 87    &       &       \cr
\hline
2   & High    & 136 & 12                     & 1460   & 146   & 2179  &       &       &       &       &       \cr
\hline
2   & Low     & 80  & 14                     & 900    & 90    & 1048  & 274   & 115   & 38    &       &       \cr
\hline
2   & High    & 150 & 13                     & 1600   & 160   & 1066  &       &       &       &       &       \cr
\hline
2   & Low     & 104 & 19                     & 1140   & 114   & 1032  & 50    & 48    & 28    &       &       \cr
\hline
2   & Low     & 112 & 20                     & 1220   & 122   & 2250  & 45    &       &       &       &       \cr
\hline
2   & Low     & 136 & 24                     & 1460   & 146   & 1981  &       &       &       &       &       \cr
\hline
2   & Low     & 150 & 27                     & 1600   & 160   & 2238  &       &       &       &       &       \cr
\hline
\end{tabular}
\end{table}

\begin{table}
\caption{\label{conf2} As table {\protect \ref{conf}}, but for three dimensions. The SU(2) data is again taken from \cite{Maas:2008ri}.}
\vspace{1mm}
\begin{tabular}{|c|c|c|c|c|c|c|c|c|c|c|c|}
\hline
$d$ & $\beta$ & $N$ & $V^{\frac{1}{d}}$ [fm] & Ther.\ & Swe.\ & SU(2) & SU(3) & SU(4) & SU(5) & SU(6) & G$_2$ \cr
\hline
3   & High    & 4   & 0.36                   & 240    & 24    & 1070  & 982   & 832   & 574   & 465   & 574   \cr
\hline
3   & Low     & 4   & 0.72                   & 240    & 24    & 1070  & 975   & 771   & 574   & 405   & 574   \cr
\hline
3   & High    & 8   & 0.72                   & 280    & 28    & 1133  & 967   & 707   & 684   & 230   & 520   \cr
\hline
3   & High    & 12  & 1.1                    & 320    & 32    & 1128  & 932   & 730   & 624   & 410   & 531   \cr
\hline
3   & Low     & 8   & 1.4                    & 280    & 28    & 1070  & 967   & 701   & 684   & 460   & 596   \cr
\hline
3   & High    & 16  & 1.4                    & 360    & 36    & 1144  & 982   & 774   & 572   & 435   & 648   \cr
\hline
3   & High    & 20  & 1.8                    & 400    & 40    & 1100  & 931   & 706   & 570   & 46    & 285   \cr
\hline
3   & Low     & 12  & 2.2                    & 320    & 32    & 1128  & 930   & 724   & 570   & 422   & 507   \cr
\hline
3   & High    & 24  & 2.2                    & 440    & 44    & 1084  & 939   & 970   & 307   &       & 68    \cr
\hline
3   & Low     & 16  & 2.9                    & 360    & 36    & 1048  & 985   & 786   & 572   & 229   & 616   \cr
\hline
3   & High    & 32  & 2.9                    & 520    & 52    & 960   & 935   & 270   & 60    &       &       \cr
\hline
3   & Low     & 20  & 3.6                    & 400    & 40    & 1217  & 979   & 734   & 617   & 33    & 172   \cr
\hline
3   & High    & 40  & 3.6                    & 600    & 60    & 1065  & 102   & 74    &       &       &       \cr
\hline
3   & Low     & 24  & 4.3                    & 440    & 44    & 1238  & 867   & 719   & 276   &       & 35    \cr
\hline
3   & High    & 48  & 4.3                    & 680    & 68    & 1095  & 100   &       &       &       &       \cr
\hline
3   & High    & 56  & 5.0                    & 760    & 76    & 1078  & 16    &       &       &       &       \cr
\hline
3   & Low     & 32  & 5.8                    & 520    & 52    & 899   & 665   & 193   & 62    &       &       \cr
\hline
3   & High    & 64  & 5.8                    & 840    & 84    & 1062  &       &       &       &       &       \cr
\hline
3   & Low     & 40  & 7.2                    & 600    & 60    & 1040  & 156   & 75    &       &       &       \cr
\hline
3   & Low     & 48  & 8.6                    & 680    & 68    & 1047  & 74    &       &       &       &       \cr
\hline
3   & Low     & 56  & 10                     & 760    & 76    & 1053  & 17    &       &       &       &       \cr
\hline
3   & Low     & 64  & 12                     & 840    & 84    & 1033  &       &       &       &       &       \cr
\hline
\end{tabular}
\end{table}

Configurations are then obtained by a mix of heat-bath and overrelaxation sweeps. This has been done for SU(2) as described in \cite{Cucchieri:2006tf}. For SU($N>3$), heat-bath updates have been performed using the Cabibbo-Marinari method \cite{Cabibbo:1982zn} with 3, 6, 10, and 15 SU(2) subgroups for $N=3$, 4, 5, and 6, respectively, according to \cite{Maas:2007af}. Between two heat-bath sweeps five overrelaxation sweeps have been performed. In the case of G$_2$, heat-bath updates have been performed according to the method presented in \cite{Pepe:2006er} and detailed in \cite{Greensite:2006sm}, and overrelaxation sweeps according to \cite{Maas:2007af}. This has been done in both, two and three dimensions. The number of configurations are given in table \ref{conf} and \ref{conf2} for two and three dimensions, respectively.

Once an equilibrated configuration is obtained, it is still necessary to fix it to the Landau gauge. This is done by minimizing the functional (see, e.g., \cite{Cucchieri:1995pn})
\be
{\cal E}=-\sum_{x,\mu}\Re\tr U_\mu,\label{gfc}
\ee
\no where $\{ U_{\mu}(x) \}$ is a thermalized lattice configuration, and the sum is over all links. This leads to the so-called minimal \cite{Cucchieri:1995pn}, or average-$B$ \cite{Maas:2009se}, Landau gauge. In principle, for comparison to the results of section \ref{sdse}, it is necessary to fix to the non-perturbative version of the Landau gauge which exhibits scaling, if such a gauge indeed exists \cite{Maas:2010wb}. For the volumes available here for SU($N>2$), this appears to be just a quantitative effect in the SU(2) case for the actual propagators \cite{Maas:2008ri,Maas:2009se}, and only slightly modifies the volume dependence. For the purpose here, investigating the changes with the gauge group, these effects are sub-leading at the present volumes, and therefore the much cheaper minimal Landau gauge is sufficient. However, it is not known yet, whether the differences between the various Landau gauges is stronger for gauge groups different from SU(2). Studies in four dimensions with SU(3) do not suggest so \cite{Bornyakov:2009ug}. Hence, this will be assumed henceforth. In particular, it will be assumed that a qualitatively similar result for the gauge-group dependence in minimal Landau gauge for the volumes investigated here implies the same qualitative behavior in other Landau gauges as well.

It is then sufficient to fix to the minimal Landau gauge. For SU(2), this is done using stochastic overrelaxation with adaptive parameter adjustment \cite{Cucchieri:2006tf}. For the other gauge groups, only the standard overrelaxation step has to be modified compared to the one in SU(2). This can be done by overrelaxing all SU(2) subgroups, in the same manner as for overrelaxation sweeps during the generation of configurations, in particular the same number of subgroups. For SU(3) the procedure is described in \cite{Suman:1993mg}, and can straightforwardly be generalized to higher values of $N$. For G$_2$ this is described in \cite{Maas:2007af}. The quality of the gauge-fixing is monitored using the quantity $e_6$, defined as \cite{Cucchieri:1995pn}
\bea
e_6&=&\frac{1}{d}\sum_\mu\frac{1}{N}\sum_c\frac{1}{[\tr(Q_\mu t_c)]^2}\times\sum_{x_\mu}(\tr\{[q_\mu(x_\mu)-Q_\mu]\tau_c\})^2\label{e6}\\
q_\mu(x_\mu)&=&\frac{1}{2i}\sum_{x_\nu,\nu\neq\mu}\big[g(x)U_\mu(x)g(x+e_\mu)^+-g(x+e_\mu)U_\mu(x)^+ g(x)^+ \big] \nn \\
Q_\mu&=&\frac{1}{N_\mu}\sum_{x_\mu}q_\mu(x_\mu)\nn,
\eea
\no where $\{ g (x) \}$ represents the gauge transformation applied on the link variables $ U_{\mu}(x) $, the symbol $+$ indicates Hermitian conjugation, $ N $ is the lattice side of the symmetric hypercube, $d$ is the space-time dimensionality, $e_{\mu}$ is a positive unit vector in the $\mu$ direction and $\tau_{c}$ are the generators of the algebra. This quantity is a more reliable measure of the gauge-fixing quality than just the transversality itself \cite{Cucchieri:1995pn}. Furthermore, it is found that the same limit of $e_6$ corresponds to a much better fulfillment of the transversality condition with increasing number of generators. Therefore, the restriction on $e_6$ for achieving the gauge-fixing can be taken somewhat lower for G$_2$ and SU($N>2$) than for SU(2), corresponding still to a better level of transversality on the average. The limits adopted here are 10$^{-12}$ for SU(2) and 10$^{-11}$ otherwise.

\section{Propagators}\label{sprop}

The determination of the propagators proceeds in the same way for all gauge groups. 

The gluon propagator is given by the correlation function 
\be
D_\mn^{ab}(p)=\frac{1}{V}<A_\mu^a(p)A_\nu^b(-p)>,\label{gpbas}
\ee
\no with the momentum-space lattice gluon field defined as 
\be
A_\mu^a(p)=e^{- \frac{i\pi p_{\mu}}{N}} \sum_x \frac{e^{2\pi i px/N}}{4i}\tr\left[\left(U_\mu(x)-U_\mu(x)^{+}\right)\tau_a\right] .\label{eq:Aofp}
\ee
\no Here the components $p_\mu$ of $p$ have the integer values $-N/2 + 1\, , \ldots , \, N/2 \, $. 

After contracting \pref{gpbas} with a transverse projector and a unit matrix in color space, the scalar part of the gluon propagator is given by
\be
D(p)=\frac{1}{V \, {\cal N}}\sum_{\mu,a}<\left[\Re A_\mu^a(p)\right]^2+\left[\Im A_\mu^a(p)\right]^2> ,\label{eq:Dofp}
\ee
\no where $\Re A_\mu^a(p)$ and $\Im A_\mu^a(p)$ are, respectively, the real and the imaginary part of $A_\mu^a(p)$ and the normalization ${\cal N}$ is given by $dN_g$ for $p>0$ and by $(d-1)N_g$ for $p=0$. $N_g$ is the number of generators and thus the number of gluons, for a given gauge group. Hence, $N_g=N^2-1$ for SU($N$) and $N_g=14$ for G$_2$. The gluon propagator is thereby by definition positive semi-definite.

The components of the physical momenta are given by
\be
P_\mu=2\sin\frac{\pi p_\mu}{N_\mu}.
\ee
\no Results will be presented as a function of the magnitude of the physical momentum $p = | P | /a$ (in GeV). Note that the continuum gluon propagator is obtained by the product $\beta \, a^{2} D(k)$, since the lattice quantity $\sqrt{\beta/a^{d-2}} A^a_{\mu}(x)$ yields the continuum quantity $A^a_{\mu}(x)$ in the formal continuum limit $a \to 0$. In the same limit, $\sqrt{\beta \, a^{d+2}} A^a_{\mu}(p)$ converges to the continuum momentum-space gluon field $A^a_{\mu}(p)$. Thus, for any dimension $d$, the lattice quantity $\beta \, a^{2} D(k)$ converges to the continuum gluon propagator in momentum space, independently of the gauge group.

The ghost propagator is given by
\be
D_G^{ab}(p)=\frac{1}{V}< (M^{-1})^{ab}(p) >, \label{eq:DofG}
\ee
\no where $M^{ab}(x,y)$ is the Faddeev-Popov operator, defined in the continuum as
\be
-\pdm D^{ab}_\mu=\delta(x-y) (-\pd^2\delta^{ab}+gf^{abc}\pd_\mu A_\mu^c) \, .
\ee
\no On the lattice in Landau gauge this operator is a matrix defined by its action on a scalar function $\omega^b(x)$ as \cite{Zwanziger:1993dh}
\bea
M(y,x)^{ab}\omega_b(x)&=&c\left(\sum_x\big(G^{ab}(x)\omega_b(x)+\sum_\mu A_\mu^{ab}(x)\omega_b(x+e_\mu)+B_\mu^{ab}(x)\omega_b(x-e_\mu)\big)\right)\nn\\
G^{ab}(x)&=&\sum_\mu \tr(\{\tau^a,\tau^b\}(U_\mu(x)+U_\mu(x-e_\mu)))\nn\\
A_\mu^{ab}(x)&=&-2\tr(\tau^a \tau^bU_\mu(x))\nn\\
B_\mu^{ab}(x)&=&-2\tr(\tau^a \tau^bU_\mu^{+}(x-e_\mu))\nn,
\eea
\no where $c$ is a constant depending on the normalization of the generators of the gauge algebra.

The evaluation of the Fourier transform of the inverse operator
\be
(M^{-1})^{ab}(p,q)=\sum_{x,y} e^{2\pi i (px+qy)/N}(M^{-1})^{ab}(x,y)
\ee
\no with $p=-q$ requires a matrix inversion, which has been performed for all gauge groups using the point source $\delta^{ac}(\delta_{x0}-1/N^d)$ \cite{Boucaud:2005gg}. Independent of the gauge group the Faddeev-Popov operator $M^{ab}(x,y)$ is symmetric and positive. Hence, in practice the matrix inversion has been performed using a conjugate-gradient method \cite{Cucchieri:2006tf}. Fortunately, the larger the number of generators the smaller is the extensive statistical noise induced by using a point source instead of a plain wave source. However, better accuracy is required in the inversion process such that the spectrum of the Faddeev-Popov operator is positive with an increasing number of generators. Otherwise negative eigenvalues are encountered as numerical artifacts, which become positive when increasing the accuracy \cite{Meurant:2006}. This procedure is ambiguous with respect to the sign of the resulting propagator (or of the eigenvalues) \cite{Cucchieri:2006tf}. Thus, the sign has to be assigned by hand, and is fixed by comparing the propagator at large momenta to perturbation theory. 

Finally, the color-averaged ghost propagator is then defined as
\be
D_G(p)=\frac{1}{N_g}D^{aa}_G.\label{cavgh}
\ee
\no One particular question is whether it is sufficient to investigate the color-averaged propagators \pref{eq:Dofp} and \pref{cavgh}. For SU(2), this is the case \cite{Maas:2007uv,Cucchieri:2006tf}, and is not surprising due to the vanishing of the symmetric structure constants. For other gauge groups, there is no reason why this should be the case. In the next section, it will be shown that this justified.

\section{Results}\label{sresults}

\subsection{Propagators}

\begin{figure}[t!]
\includegraphics[width=\linewidth]{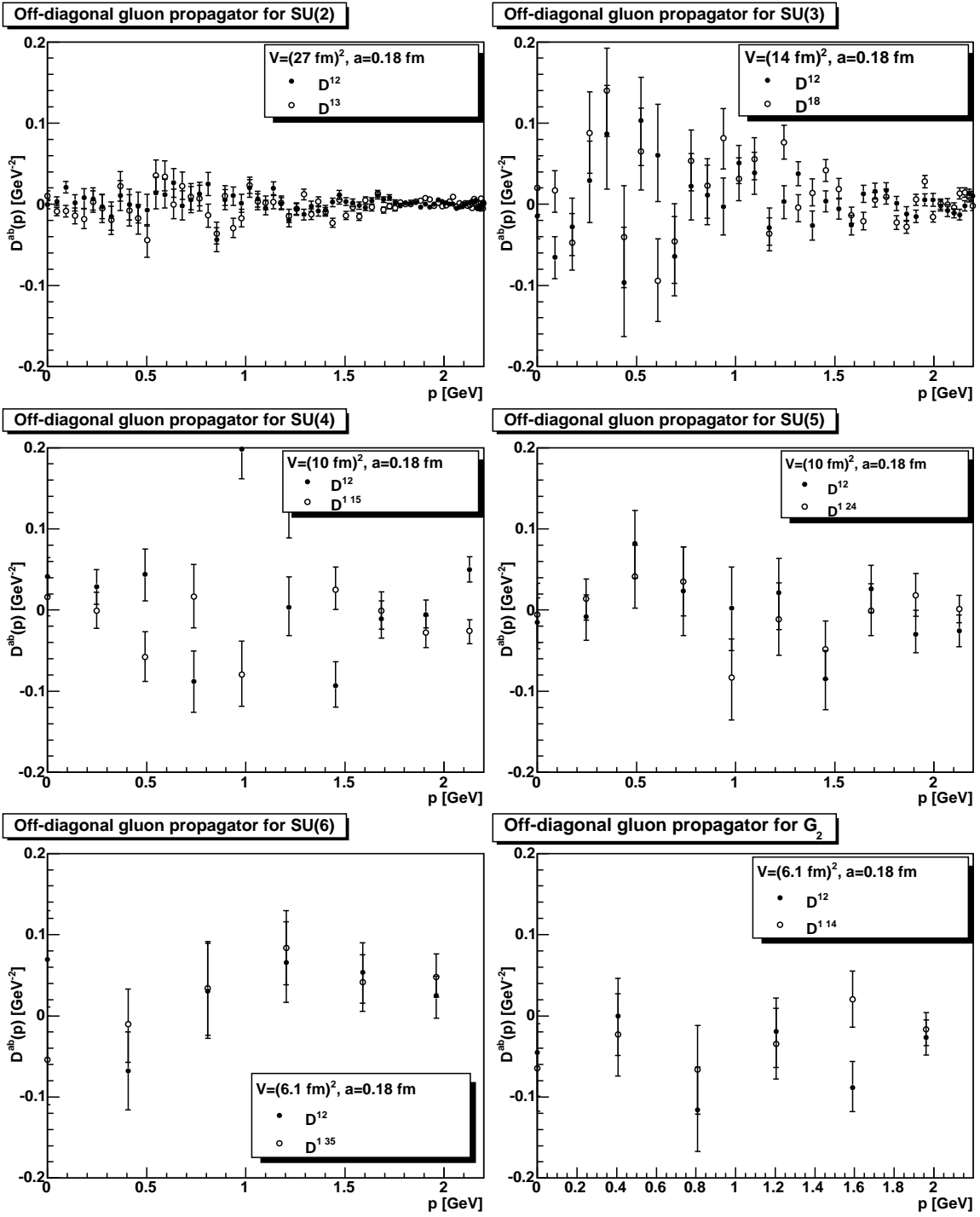}
\caption{\label{gp-off-2d}The off-diagonal gluon propagator in two dimensions for the different gauge groups. Always the components $D^{12}$ (full symbols) and $D^{1N_g}$ (open symbols) are shown.}
\end{figure}

\begin{figure}
\includegraphics[width=\linewidth]{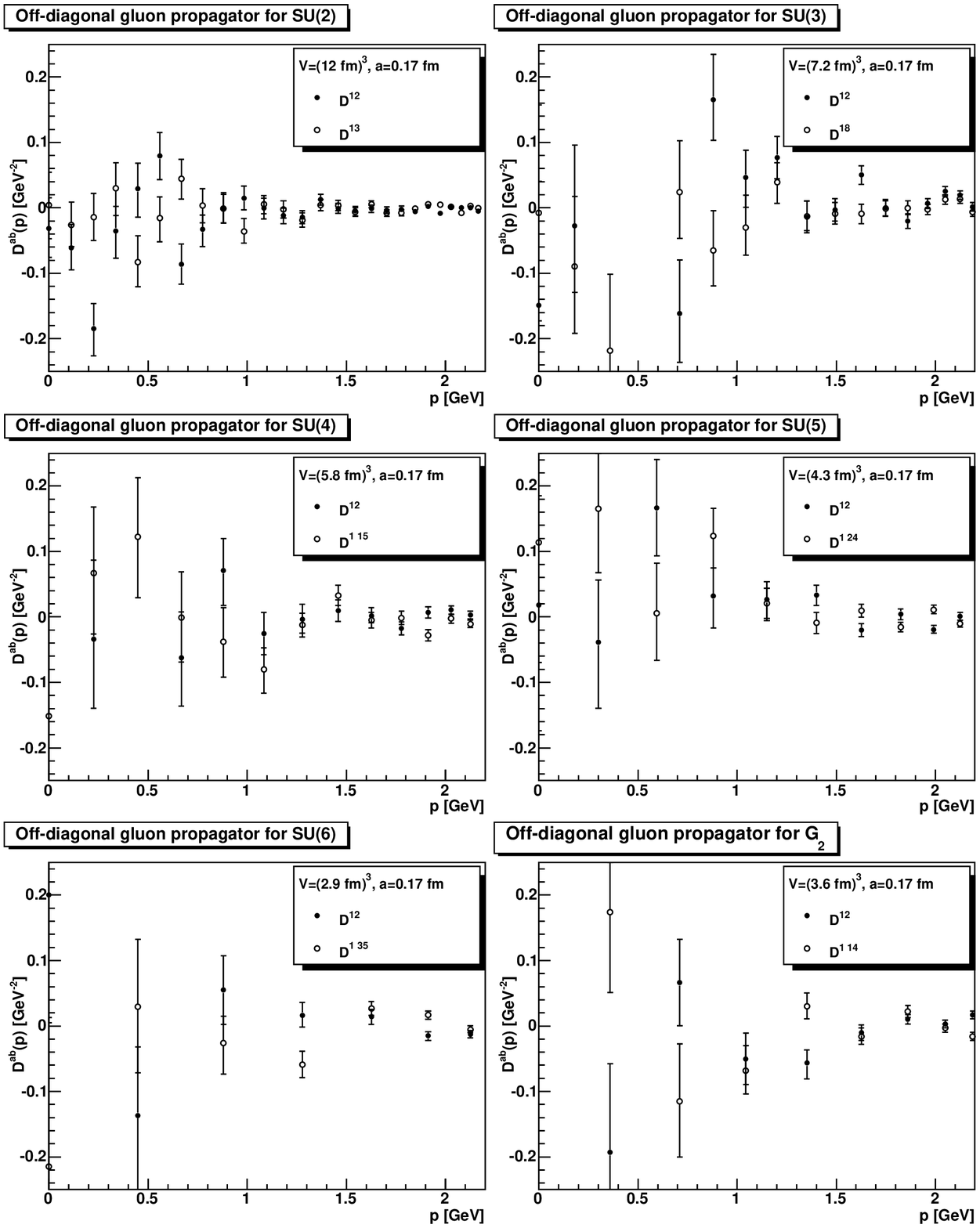}
\caption{\label{gp-off-3d}The off-diagonal gluon propagator in three dimensions for the different gauge groups. Always the components $D^{12}$ (full symbols) and $D^{1N_g}$ (open symbols) are shown.}
\end{figure}

The initial check is whether the color-off-diagonal propagator elements indeed vanish, as this was a necessary precondition for the analysis of section \ref{sdse}, and is a general assumption in functional studies. The first results are therefore for the color off-diagonal, but Lorentz-structure averaged, gluon propagator in figure \ref{gp-off-2d} in two dimensions and in figure \ref{gp-off-3d} for three dimensions, for the various gauge groups. In both two and three dimensions and for all gauge groups the results are, within statistical errors, in agreement with a zero result. The gluon propagator is thus color-diagonal, at the very least for the lattice settings studied here.

\begin{figure}
\includegraphics[width=\linewidth]{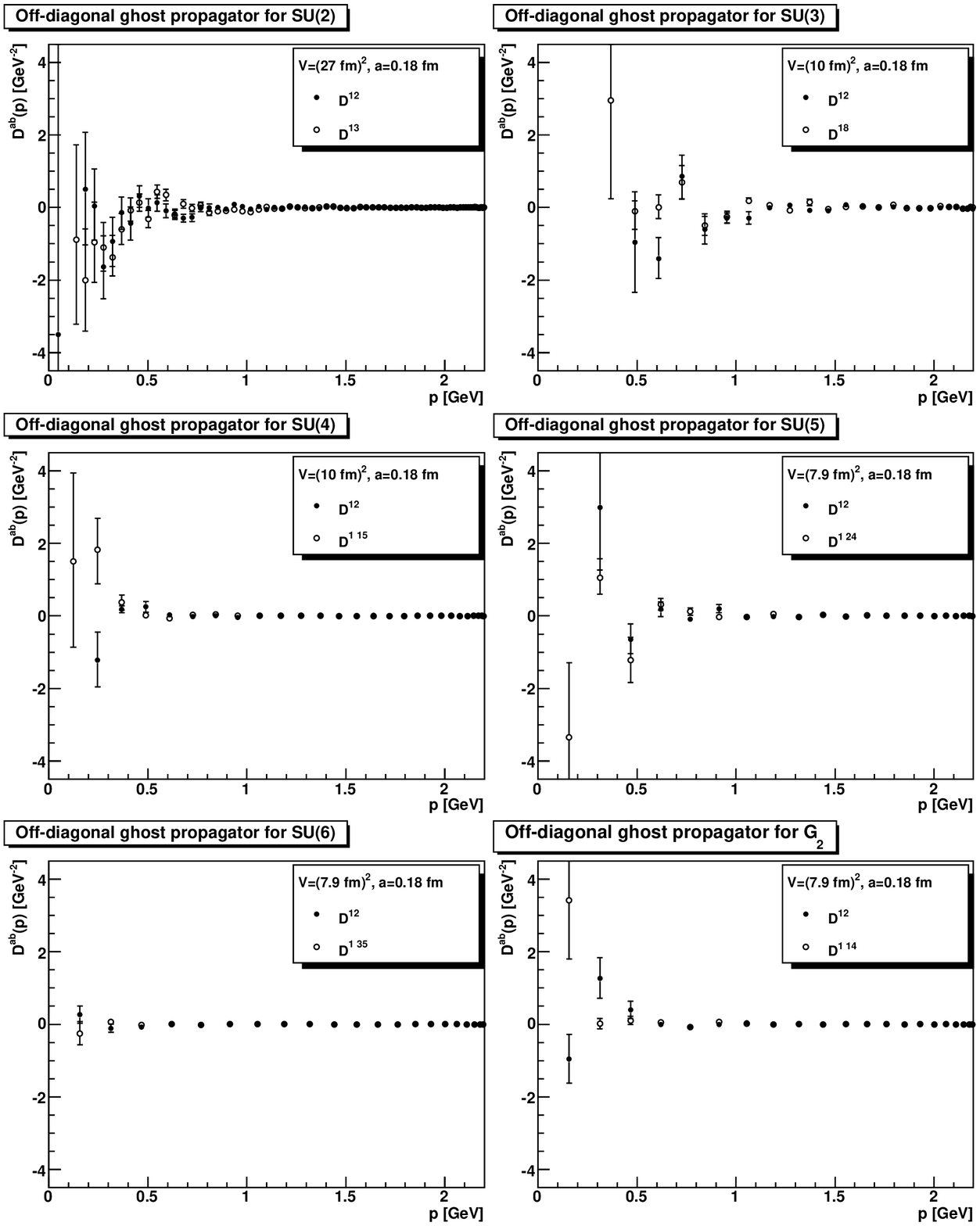}
\caption{\label{ghp-off-2d}Same as in figure {\protect\ref{gp-off-2d}}, but for the ghost propagator.}
\end{figure}

\begin{figure}
\includegraphics[width=\linewidth]{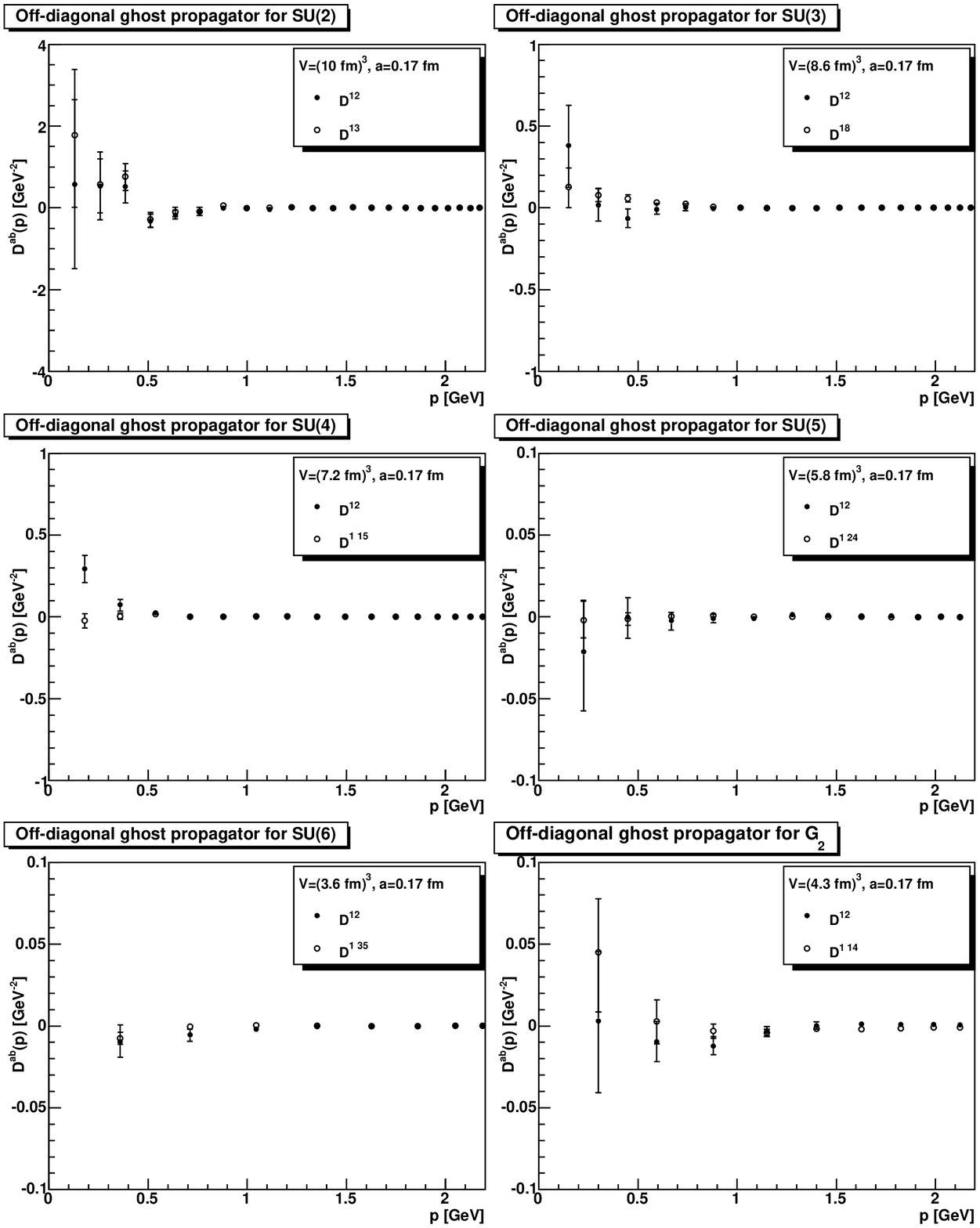}
\caption{\label{ghp-off-3d}Same as in figure {\protect\ref{gp-off-3d}}, but for the ghost propagator.}
\end{figure}

The corresponding result for the ghost propagator is shown in figure \ref{ghp-off-2d} for two dimensions and in figure \ref{ghp-off-3d} for three dimensions. Once more, for all gauge groups the off-diagonal elements are consistent with zero, in agreement with the arguments of section \ref{sdse}.

This supports corresponding assumptions in functional calculations of all types. Note that this is not necessarily implying a simple color-structure on the level of the vertices \cite{Alkofer:2008dt,Kellermann:2008iw}. To some extent, this is an expected result, as off-diagonal color elements could be indicative of the breaking of the residual global color symmetry, which would be unexpected. However, note that it is currently unclear whether in the minimal Landau gauge the corresponding charge can be well-defined for a photon-ghost behavior \cite{Kugo:1979gm}.

\begin{figure}[htpb]
\includegraphics[width=0.95\linewidth]{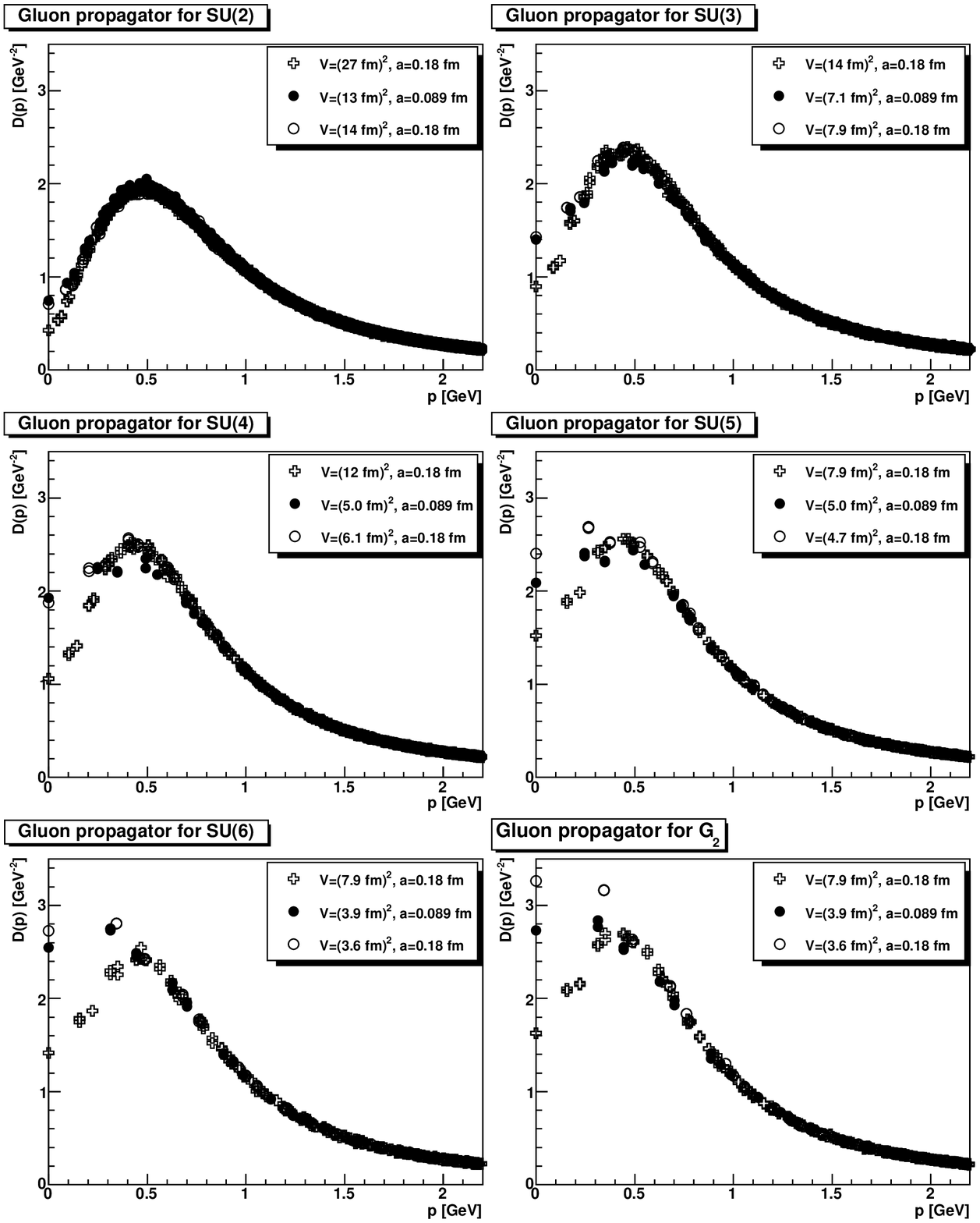}
\caption{\label{gp-2d}The gluon propagator in two dimensions for the various gauge groups. Error bars are partly smaller than the symbol size. Various momentum configurations are shown, see {\protect \cite{Cucchieri:2006tf}} for details.}
\end{figure}

\begin{figure}
\includegraphics[width=\linewidth]{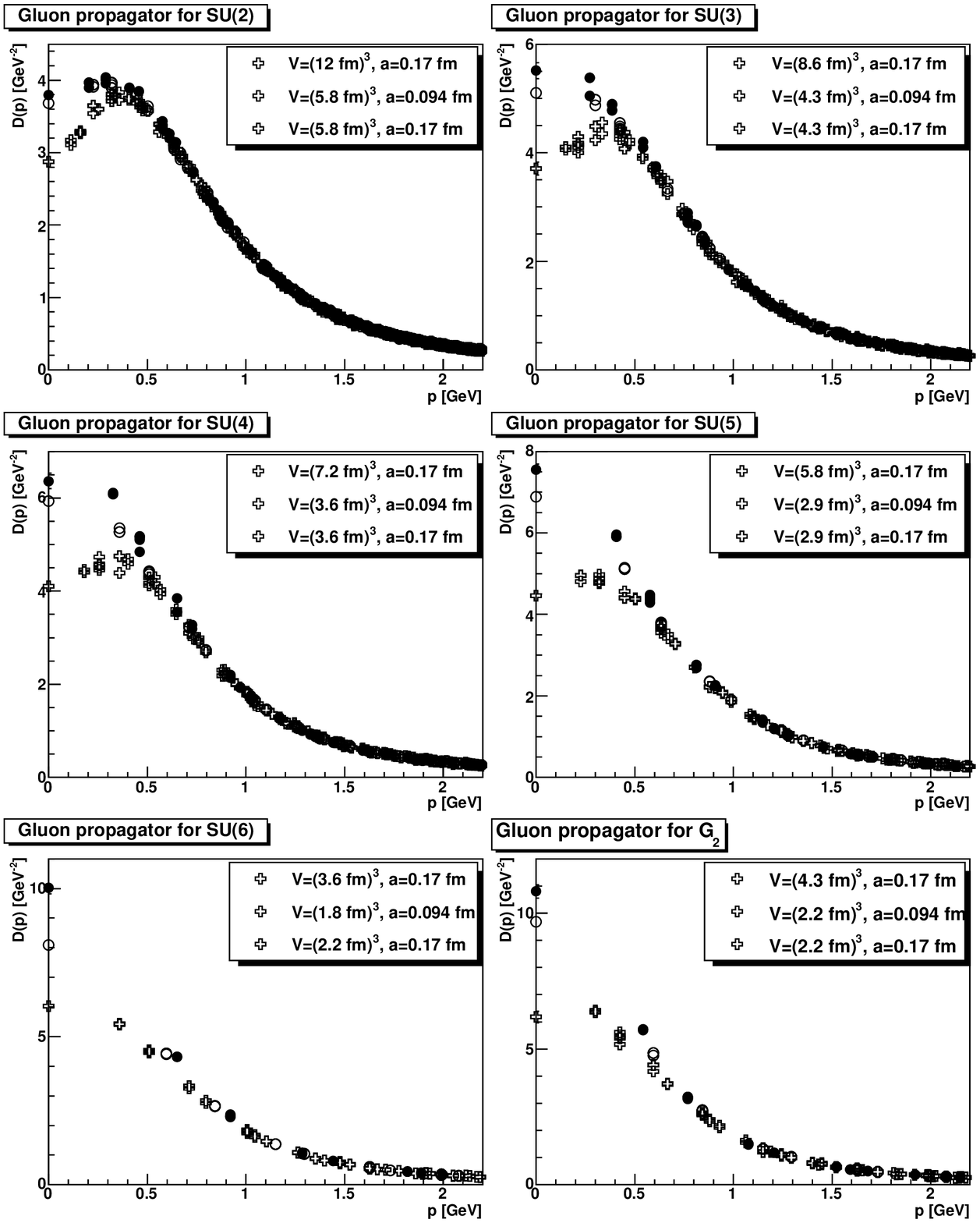}
\caption{\label{gp-3d}Same as in figure {\protect \ref{gp-2d}}, but for three dimensions.}
\end{figure}

The results for the color-diagonal gluon propagator is shown in figure \ref{gp-2d} for two dimensions and in figure \ref{gp-3d} for three dimensions.

In two dimensions, for all gauge groups, the gluon propagator is infrared suppressed, compared to tree-level. In fact, for sufficiently large volumes it is also stronger suppressed than the one of a massive particle, as expected from section \ref{sdse} and the previous results for SU(2) alone \cite{Cucchieri:2007rg,Maas:2007uv}. Furthermore, the relevant scales, expressed in units of the string tension, turn out to be essentially the same for all gauge groups: The maximum occurs at about half a GeV, and the height of the maximum is between two and three inverse GeV$^2$. Also, in all cases the effects of the violation of rotational symmetry are rather small. The infrared suppression observed is in all cases also sufficient to cure the infrared problems encountered in perturbation theory, making all integrals in the Dyson-Schwinger equations, or elsewhere, well-defined.

In three dimensions, only in the SU(6) (and possibly the G$_2$) case an explicit maximum is not visible, although the finite-volume effects for all gauge groups demonstrate explicitly that $D(0)$ has still not settled on its infinite-volume value. Hence, the existence of a maximum at larger volumes for the group SU(6) appears likely. Furthermore, for the same volume for all other gauge groups also no maximum is visible. Hence, the qualitative turn-over structure seems to agree for all gauge groups in three dimensions.

\begin{figure}
\includegraphics[width=\linewidth]{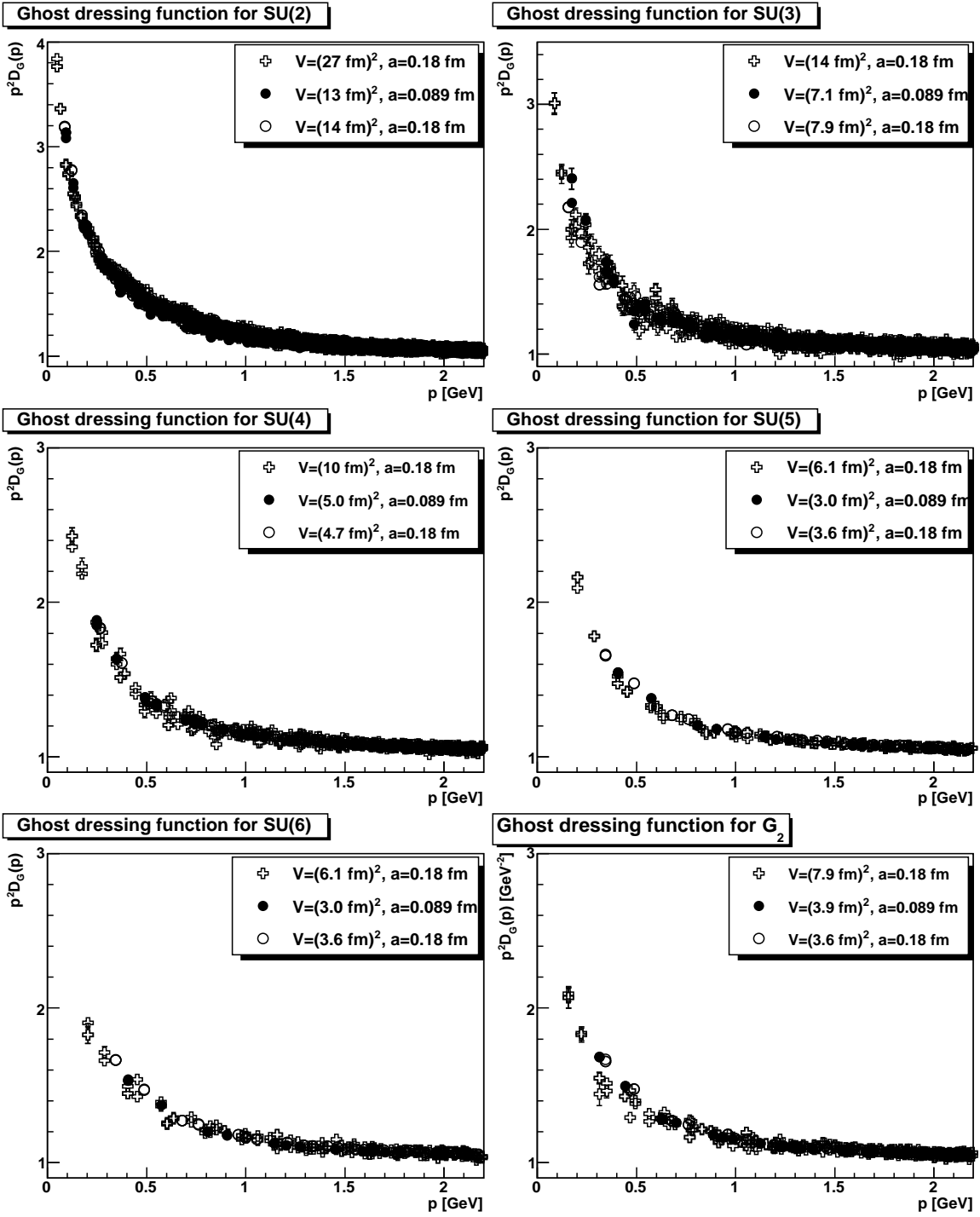}
\caption{\label{ghp-2d}The ghost dressing function for the various gauge groups in two dimensions. Error bars are partly smaller than the symbol size. Various momentum configurations are shown, see {\protect \cite{Cucchieri:2006tf}} for details.}
\end{figure}

\begin{figure}
\includegraphics[width=\linewidth]{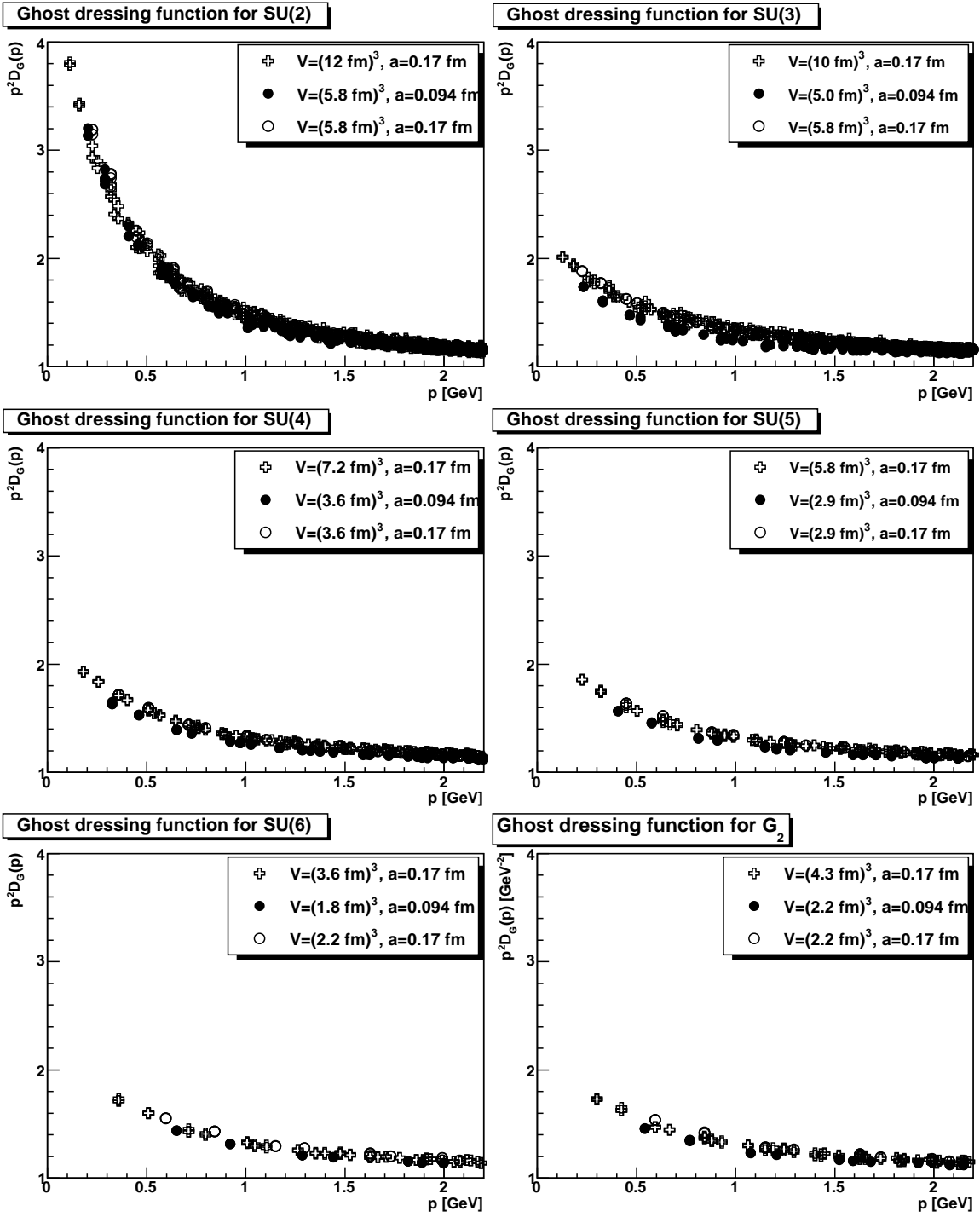}
\caption{\label{ghp-3d}Same as in figure {\protect \ref{ghp-2d}}, but for three dimensions.}
\end{figure}

The results for the ghost dressing function are shown in figure \ref{ghp-2d} for two dimensions and in figure \ref{ghp-3d} for three dimensions.

In two dimensions again there is no qualitative, and little quantitative, difference between all gauge groups. The ghost dressing function is in all cases infrared enhanced, and compatible with being infrared divergent. Also, the effects of violation of rotational invariance are once more rather small. Furthermore, the ghost dressing function is for all gauge groups monotonous for the complete momentum regime. Thus, as in case of the gluon propagator, the similarity between the various gauge groups holds also beyond the asymptotic regimes discussed in section \ref{ssce}.

In three dimensions, in all cases, the dressing function is infrared enhanced. In case of SU(2), substantial evidence exists that this enhancement is not a divergence, and the ghost dressing function becomes finite at very small momenta \cite{Cucchieri:2008fc}, despite its apparent strength. Since the propagators for the other gauge groups are similar to the one of SU(2) at equivalent volumes, though less strongly enhanced, they will likely become also finite at larger volumes. Nonetheless, the main result here is that these ghost dressing function are qualitatively very similar, and the influence of the gauge algebra on the ghost propagator at these volumes is therefore only quantitative.

\subsection{Asymptotic behavior}\label{sasymp}

\subsubsection{Ultraviolet behavior}

\begin{table}
\caption{\label{uvfit} Fit results for the ultraviolet fits. Errors are statistical only.}
\vspace{1mm}
\begin{tabular}{|c|c||c|c|c|c||c|c|c|c|c|}
\hline
Propagator & Group & $d$ & $a$ & $\frac{a}{C_A}$ & $\frac{a}{C_Ag^2}$ & $d$ & $a$ & $\frac{a}{C_A}$ & $\frac{a}{C_Ag^2}$ & $\frac{a}{a_\mathrm{LO}}$ \cr
\hline
Gluon & SU(2) & 2 & $-0.43(11)$ & -0.22 & -0.43 & 3 & $-0.59(3)$ & -0.30 & -0.26 & 1.5 \cr
\hline
Gluon & SU(3) & 2 & $-0.46(12)$ & -0.15 & -0.52 & 3 & $-0.58(6)$ & -0.19 & -0.28 & 1.6 \cr
\hline
Gluon & SU(4) & 2 & $-0.47(6)$ & -0.12 & -0.55 & 3 & $-0.56(5)$ & -0.14 & -0.28 & 1.6 \cr
\hline
Gluon & SU(5) & 2 & $-0.43(6)$ & -0.086 & -0.53 & 3 & $-0.56(4)$ & -0.11 & -0.29 & 1.7 \cr
\hline
Gluon & SU(6) & 2 & $-0.50(9)$ & -0.083 & -0.63 & 3 & $-0.54(4)$ & -0.090 & -0.28 & 1.7 \cr
\hline
Gluon & G$_2$ & 2 & $-0.40^{+10}_{-11}$ & -0.20 & -0.51 & 3 & $-0.50(7)$ & -0.25 & -0.28 & 1.6 \cr
\hline
Ghost & SU(2) & 2 & $-0.28(4)$ & -0.14 & -0.28 & 3 & $-0.29(2)$ & -0.15 & -0.13 & 2.0 \cr
\hline
Ghost & SU(3) & 2 & $-0.26(5)$ & -0.087 & -0.29 & 3 & $-0.282(8)$ & -0.0940 & -0.13 & 2.1 \cr
\hline
Ghost & SU(4) & 2 & $-0.258(6)$ & -0.0645 & -0.30 & 3 & $-0.287(3)$ & -0.0718 & -0.14 & 2.3 \cr
\hline
Ghost & SU(5) & 2 & $-0.266(2)$ & -0.0532 & -0.32 & 3 & $-0.285(2)$ & -0.0570 & -0.15 & 2.4 \cr
\hline
Ghost & SU(6) & 2 & $-0.265(1)$ & -0.0442 & -0.34 & 3 & $-0.283(2)$ & -0.0472 & -0.15 & 2.4 \cr
\hline
Ghost & G$_2$ & 2 & $-0.256(4)$ & -0.128 & -0.33 & 3 & $-0.300(2)$ & -0.15 & -0.17 & 2.7 \cr
\hline
\end{tabular}
\end{table}

The first interesting question to be established is whether the universality at large momenta obtained perturbatively in higher dimensions also holds in two dimensions. For this it is necessary to obtain the leading perturbative corrections to the tree-level value. This is done by fitting the data for both propagators above 2 GeV and for the volumes used in the figures of section \ref{sresults} and the larger $\beta$ value for momenta along the diagonal with the form
\be
p^2D(p)=\frac{1}{1+\frac{a}{p^2}}\label{uvfitform}.
\ee
\no The results of these fits are given in table \ref{uvfit}. The results for the gluon propagator are found to cluster around a value of $c=a/(g^2C_A)$ of about 1/2. The values are always negative, demonstrating the existence of a Landau pole also in two dimensions. Using instead of 2 GeV 4 GeV as the starting point for the fit results in a slight increase in the value of $a$, but at the expense of a larger statistical error. A similar results is found for the ghost propagator, though the value is now closer to 3/10. Altogether, the results show that the perturbative behavior in two dimensions follows the expected pattern of gauge algebra dependence.

To compare how well this approximates the leading-order behavior also the results for three dimensions are given and compared to the leading-order result. For the gluon propagator, the deviation is sizeable\footnote{Some of the deviations could possibly be accommodated by a tadpole correction \cite{Bloch:2003sk}, which amount to about a ten-precent effect at these values of $\beta$.}, but still not dominant. The agreement is slightly but not drastically improved when starting to fit at 4 GeV instead.

For the ghost propagator, the leading-order approximation essentially breaks down at 2 GeV, and is not much better at 4 GeV. The results, which therefore include higher-order corrections, still show an approximate scaling proportional to $g^2 C_A$ with the gauge-group, showing that even sub-leading corrections are pre-dominantly of this type, rather than having more complex dependencies on the gauge algebra.

Note that despite these differences the propagators above 2 GeV are almost completely dominated by their tree-level behavior. Even with the enhancement in the ghost case, the leading-order corrections at 2 GeV amount only to a 15-20\% effect. Without it, it is only a 5\% effect.

\subsubsection{Infrared behavior in two dimensions}

\begin{figure}
\includegraphics[width=0.5\linewidth]{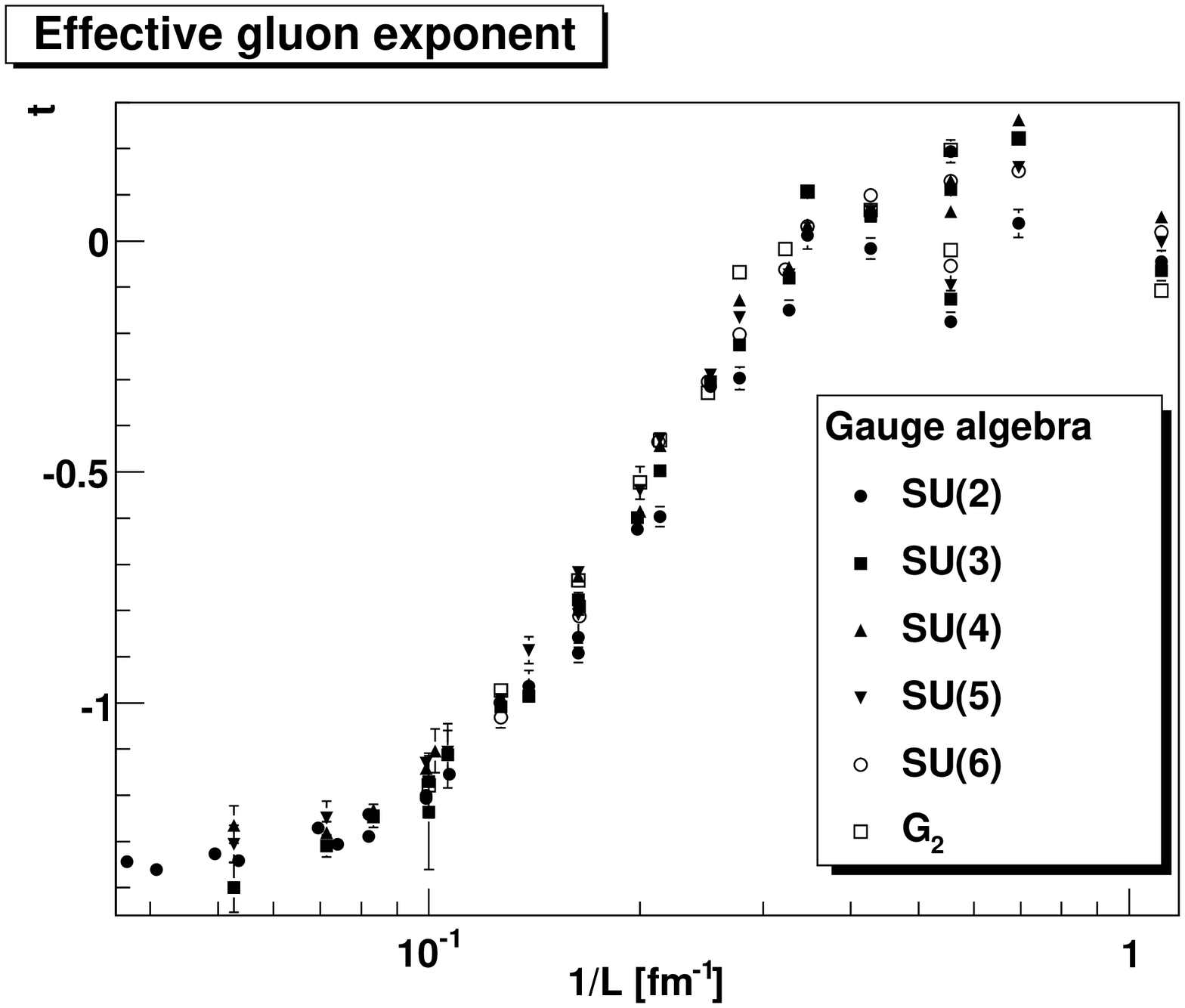}\includegraphics[width=0.5\linewidth]{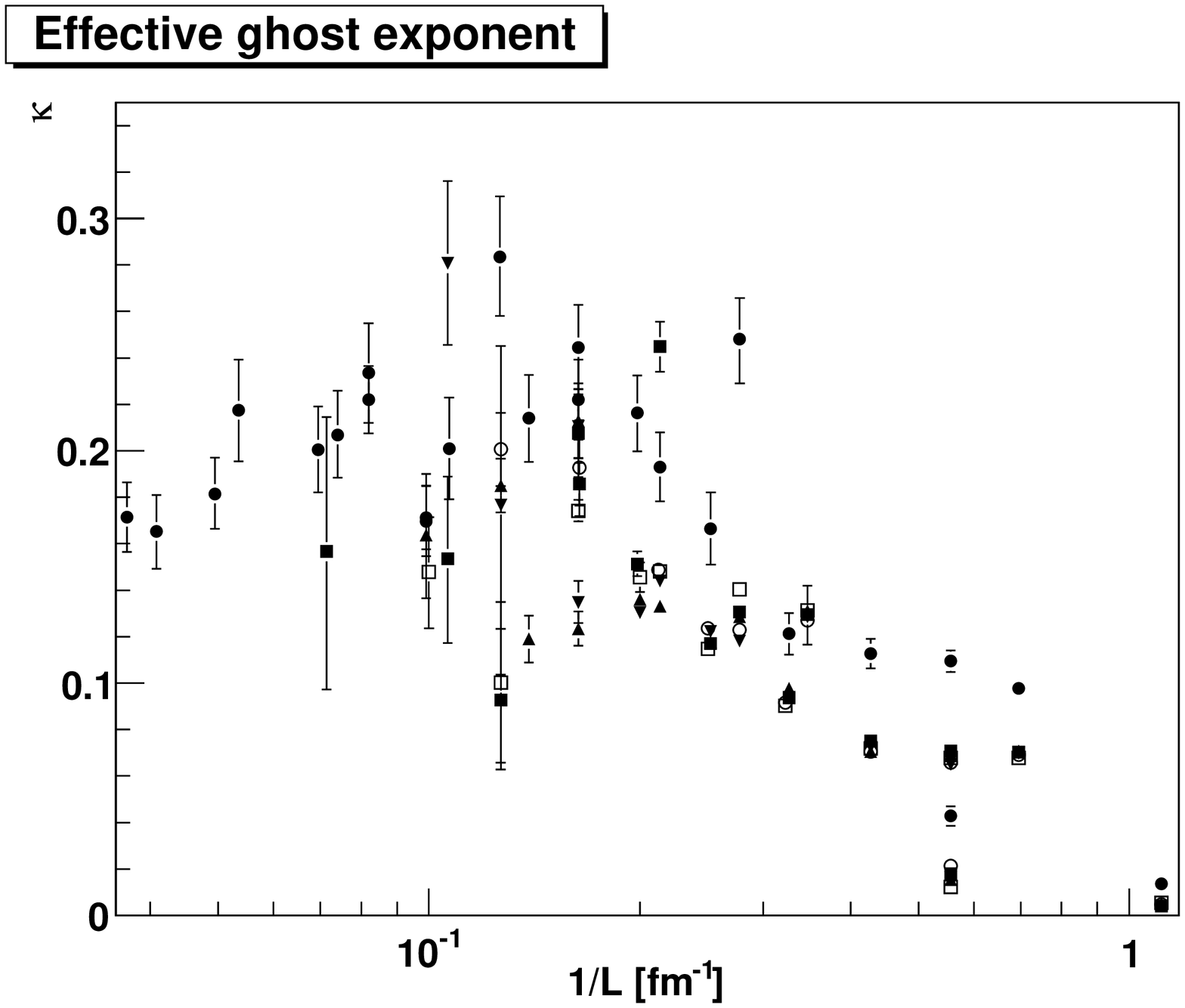}
\caption{\label{ex-2d}The effective gluon (left) and ghost (right) exponents in two dimensions. Full circles are SU(2), full squares SU(3), full triangles SU(4), full upside-down triangles SU(5), open circles SU(6) and open squares G$_2$.}
\end{figure}

As shown in figures \ref{gp-2d} and \ref{ghp-2d}, the propagators exhibit for all gauge groups a behavior which is similar to the SU(2) case in approximately similar volumes. Therefore, a fit with the ans\"atze \pref{ir1} and \pref{ir2} can be performed, as described in \cite{Maas:2007uv}. This yields volume-dependent effective exponents. They are shown in figure \ref{ex-2d} for the various gauge groups.

The curves for the gluon exponent are rather similar, and the small differences could easily be dominated by the uncertainties in the comparison of the scales. In particular, the value of the exponent at large volumes appears to be rather insensitive to the gauge group.

The same observations pertains to the ghost exponent. In general, the value of the exponent is in agreement between all gauge groups, and seems to be consistent with the same value in the infinite-volume limit. With the rather larger statistical errors, this cannot be finally settled

Note that in both cases these evolution statements can only be made under the assumption that the assigned volume scale is indeed in rough agreement between all gauge groups. Therefore, the most important statement is that the infinite-volume limit seems to be in agreement, as expected from the discussion of section \ref{sdse}. This also implies that gauge-group-dependent corrections to the low-momentum behavior of the ghost-gluon vertex are likely small and of quantitative nature only. This is in agreement with investigations of the ghost-gluon vertex for SU(2) and SU(3) in four dimensions using lattice gauge theory \cite{Cucchieri:2008qm,Ilgenfritz:2006he}, as well as with functional studies \cite{Schleifenbaum:2004id}.

\begin{figure}
\includegraphics[width=0.5\linewidth]{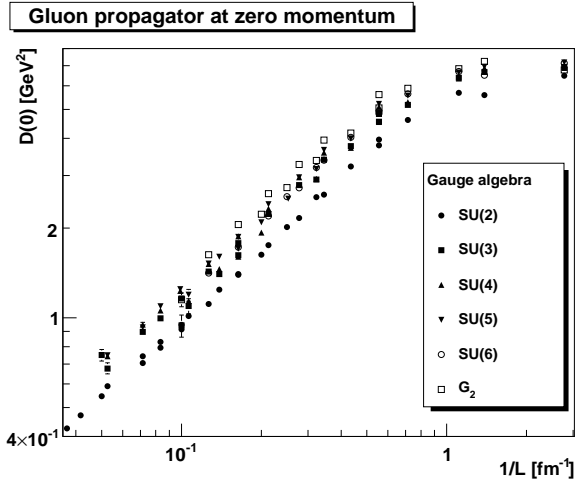}
\caption{\label{d0-2d}The gluon propagator at zero momentum in two dimensions. Full circles are SU(2), full squares SU(3), full triangles SU(4), full upside-down triangles SU(5), open circles SU(6) and open squares G$_2$.}
\end{figure}

Although not in all cases the fitted gluon exponents have yet reached a stable value as for SU(2), a maximum is already visible in all cases. If the gluon propagator is to vanish in the infinite-volume limit in two dimensions, then the value of it at zero momentum is predicted to behave as a power of inverse volume \cite{Fischer:2007pf}. This is indeed the case for all volumes investigated here, as can be seen in figure \ref{d0-2d}. In fact, for all gauge groups the gluon propagator vanishes like the same inverse power of $V$. This inverse power is of the same size as it would be expected on the basis of finite volume corrections obtained in functional calculations \cite{Maas:2007uv,Fischer:2007pf}.

\begin{figure}
\includegraphics[width=\linewidth]{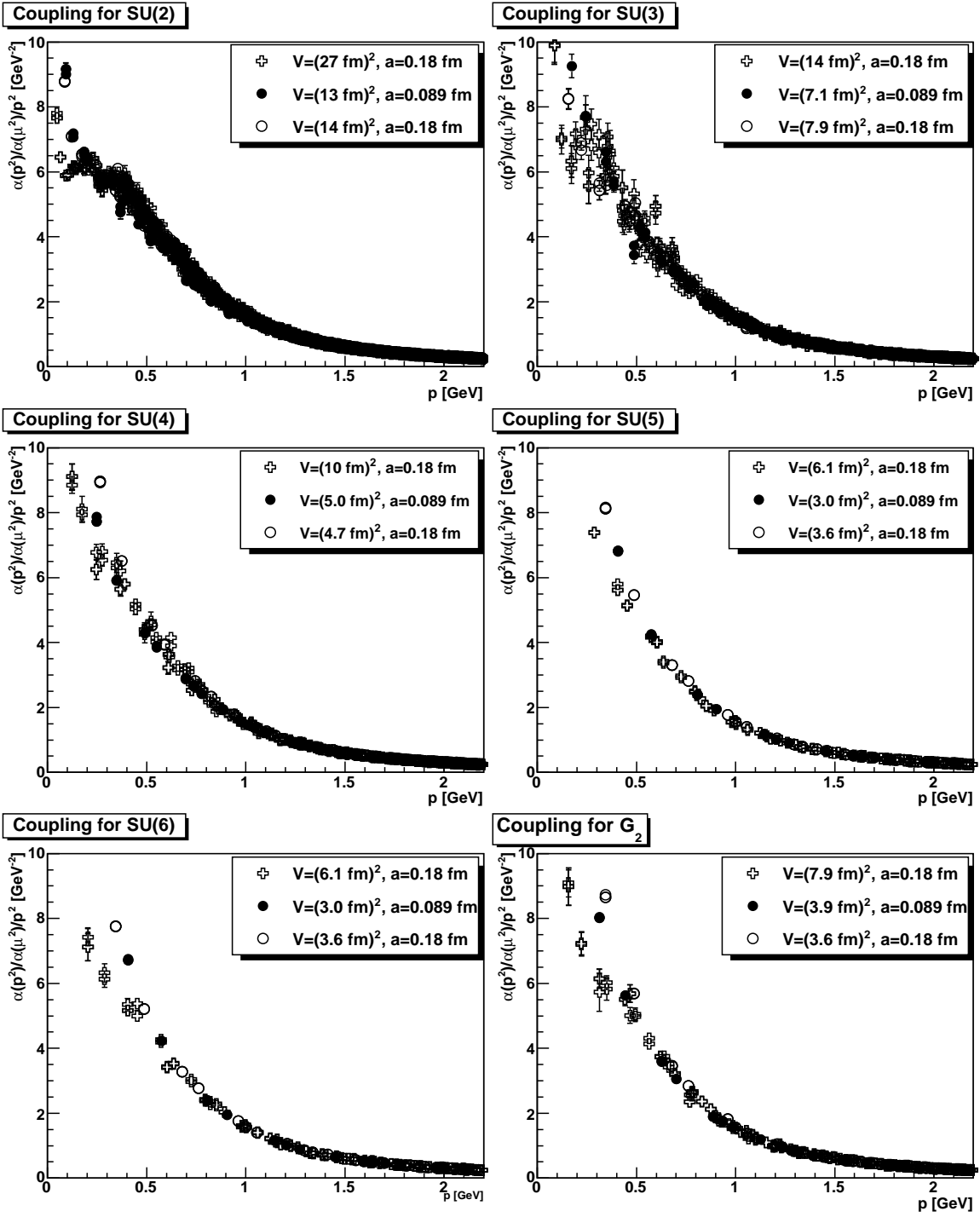}
\caption{\label{alpha-2d}The coupling {\protect \pref{coupling}} for the various gauge groups in two dimensions. Always the results for the largest volume with the low $\beta$ value (crosses) and the high $\beta$ value (full circles) are shown. In addition, to assess discretization effects, also the result for the lower $\beta$-value, which has a volume most closest to the result for the higher $\beta$ value shown, is presented as well.}
\end{figure}

The final quantity to be evaluated is the effective running coupling, which is given in \pref{coupling}. The results for the various gauge groups are shown in figure \ref{alpha-2d}. Not for all gauge groups sufficiently large volumes are available to identify the existence of an infrared plateau, as is possible for SU(2). However, the approximate value is in all cases about the same, and thus, if at all, there is only a weak dependence of the effective coupling in the infrared on the gauge group.

\subsubsection{Infrared behavior in three dimensions}

In three dimensions in minimal Landau gauge the behavior of the propagators change to a photon-ghost behavior at a momentum of order a few dozen MeV \cite{Cucchieri:2008fc,Cucchieri:2007rg}, though still at momenta much smaller than in four dimensions \cite{Bogolubsky:2009dc,Cucchieri:2008fc,Cucchieri:2007rg}. Ultimately, therefore, in minimal Landau gauge the gluon propagator resembles the one of a screened particle and the ghost propagator becomes photon-like. Still, this critical momentum is lower than accessible in most cases here. Only for SU(2), this momentum is just so reached. But since such momenta are currently almost not accessible for most gauge groups, and the intermediate low-momentum behavior appears to be power-like for all groups, this will be investigated here. Since in this intermediate momentum range a scaling-like behavior is expected \cite{Fischer:2008uz}, the analysis of section \ref{sdse} can be applied with an infrared cutoff \cite{Fischer:2007pf}. As a consequence, the behavior of the propagators in this intermediate range should again be universal, i.\ e., independent of the gauge algebra up to trivial 't Hooft scaling. This will be checked here.

\begin{figure}
\includegraphics[width=0.5\linewidth]{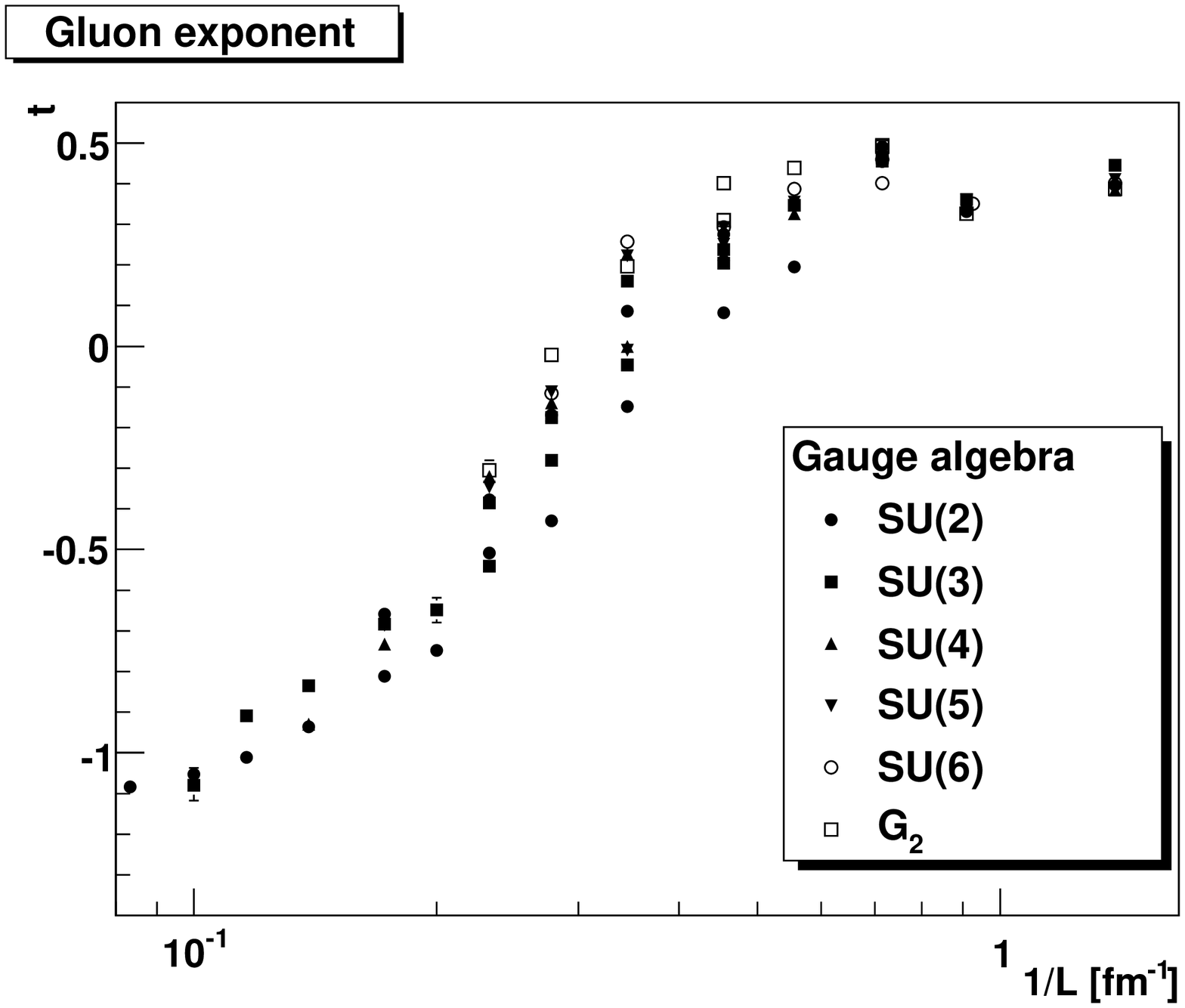}\includegraphics[width=0.5\linewidth]{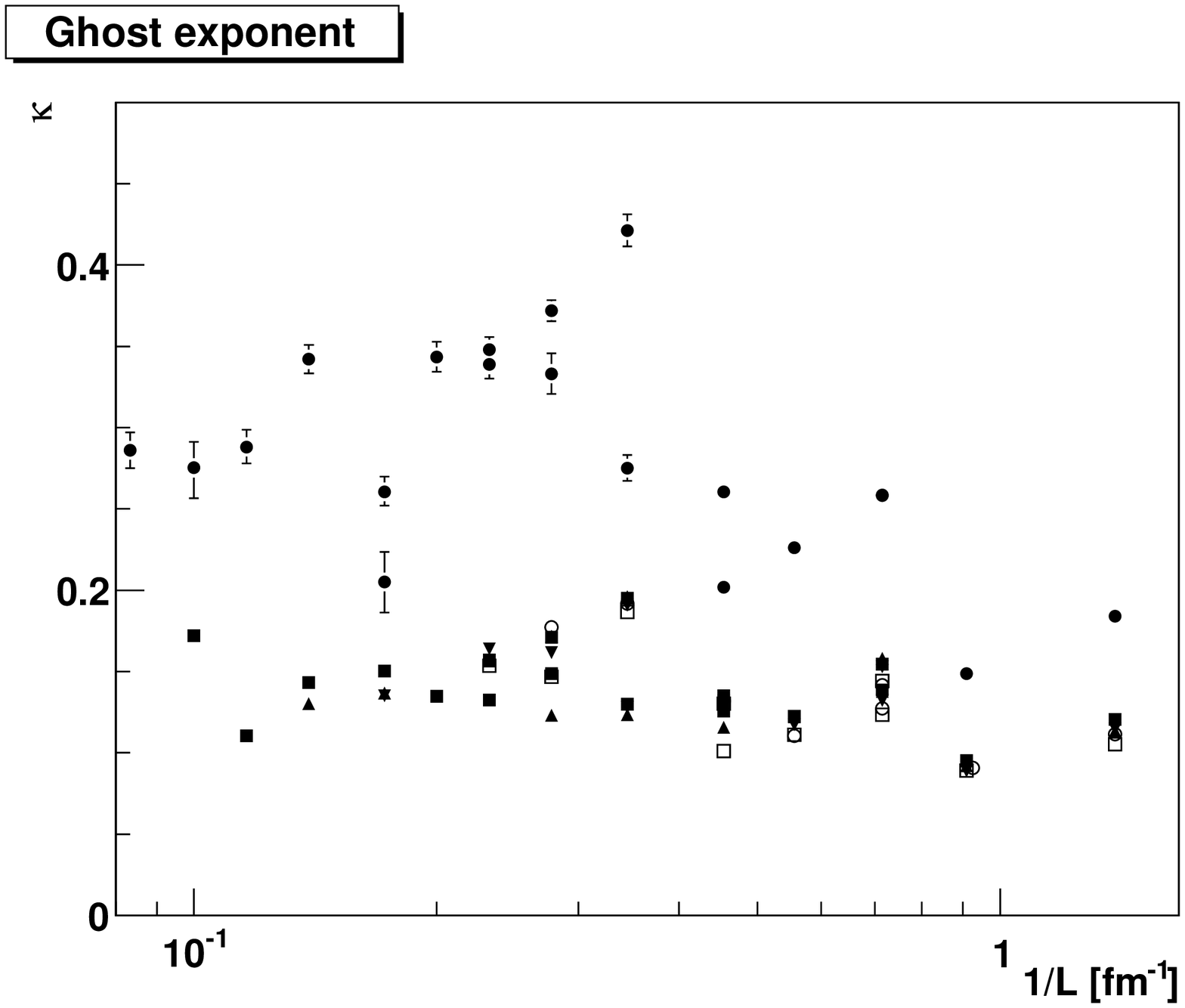}
\caption{\label{ex-3d}The effective gluon (left) and ghost (right) exponent in three dimensions, determined according to \cite{Maas:2007uv}. Full circles are SU(2), full squares SU(3), full triangles SU(4), full upside-down triangles SU(5), open circles SU(6) and open squares G$_2$. Only results with a statistical error less than 0.1 are shown for clarity.}
\end{figure}

\begin{figure}
\includegraphics[width=0.5\linewidth]{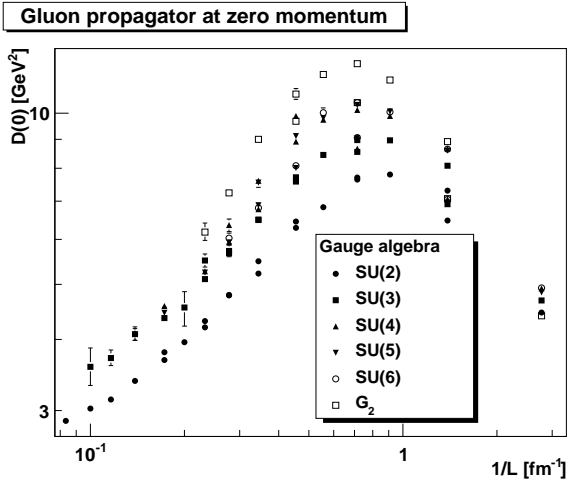}
\caption{\label{d0-3d}The gluon propagator at zero momentum in three dimensions. Full circles are SU(2), full squares SU(3), full triangles SU(4), full upside-down triangles SU(5), open circles SU(6) and open squares G$_2$.}
\end{figure}

Taking thus the low-momentum forms \pref{ir1} and \pref{ir2} as fit ans\"atze, the effective, volume-dependent exponents can be determined. These are plotted in figure \ref{ex-3d}. At least for SU(2) and SU(3) the gluon exponent becomes smaller than -1 for the given volume, i.\ e., at least for some momentum range the gluon propagator decreases faster than would be expected from a screened particle behavior. Of course, in much larger volumes this exponent increases again towards -1. Furthermore the volume-dependence of the effective exponents is once more the same for all gauge groups, as is expected from the presence of the scaling window. However, quantitatively, there is some difference in the ghost case for the gauge groups. In addition, the gluon propagator at zero momentum should decrease with volume for a certain range of volumes, like a power-law. This is indeed the case, as can be seen in figure \ref{d0-3d}.

\begin{figure}
\includegraphics[width=\linewidth]{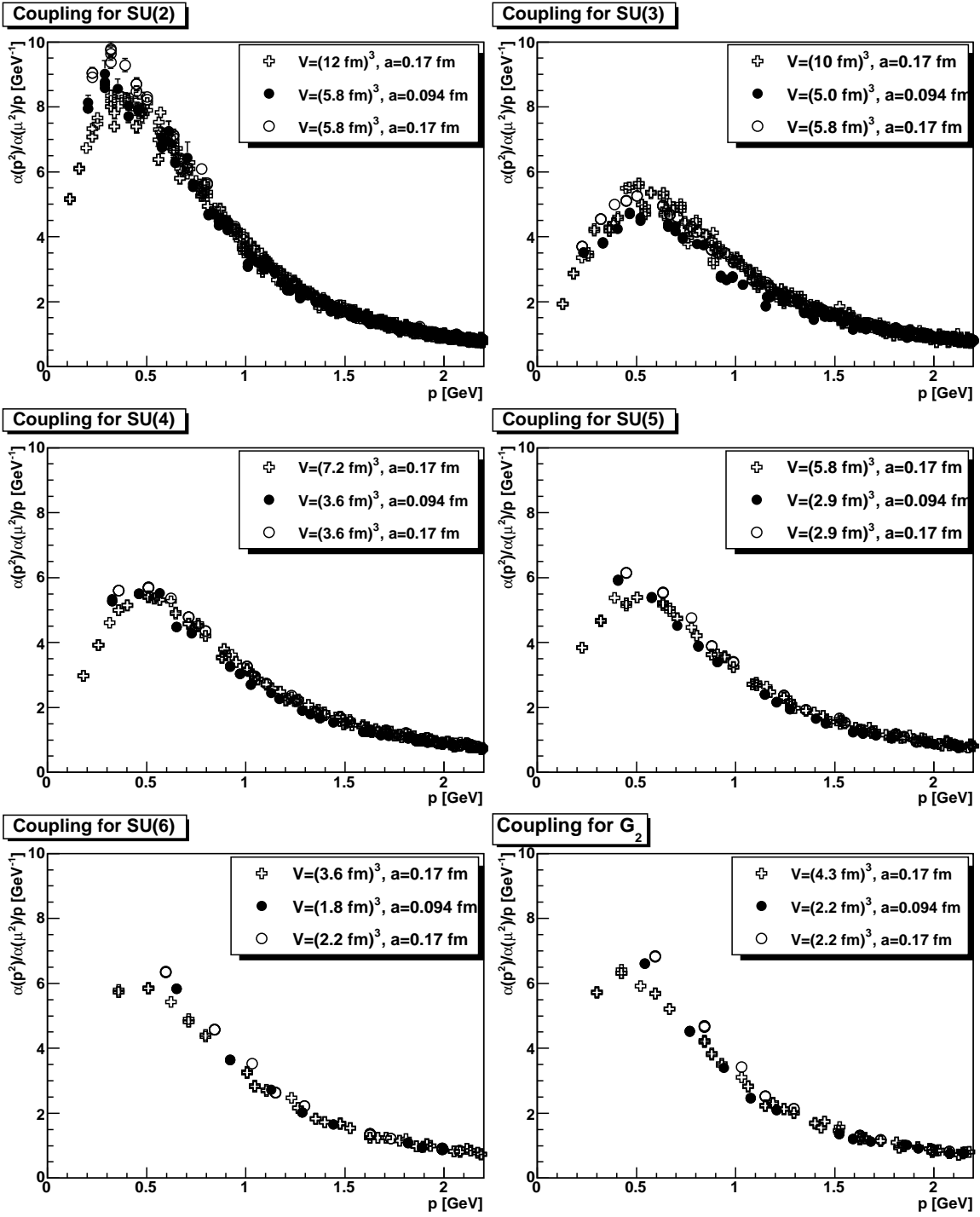}
\caption{\label{alpha-3d}Same as figure {\protect \ref{alpha-2d}}, but for three dimensions.}
\end{figure}

Finally, it is worthwhile to also investigate the effective coupling. The results are shown in figure \ref{alpha-3d}. Again, the behavior is also quantitatively rather similar. In all cases, the coupling is infrared suppressed, as is expected in any case. However, the maximum can be compared to \pref{ircoeff}. Since the highest value reached is more or less similar this implies that the maximum scales approximately like \pref{ircoeff}. The difference between SU(2) and the remaining groups maybe connected with the fact that the peak is here at about 400 instead of 500-600 MeV. There, again, some scale uncertainty may be involved.

Hence, also in three dimensions, the asymptotic behavior at low momentum follows in an intermediate volume window a scaling-type behavior, as is expected independently of the far infrared behavior. Even without investing this information, the result is that the same pattern emerges for all gauge groups, and thus even quantitatively the effects of the different gauge algebras are small. Furthermore, the behavior is in all cases not compatible with perturbation theory at low momenta, nor do non-perturbative contributions seem to diminish with $N$. Thus it is likely that even in the limit of $N\to\infty$ non-perturbative effects dominate the infrared behavior of the correlation functions, at least for momenta below $g^2 C_A$.

Of course, this is exactly as it would be expected when topological configurations, contributing with an essential singularity in the coupling constant, dominate the low-momentum behavior. There is evidence for this \cite{Maas:2008uz,Maas:2005qt,Greensite:2004ur,Gattnar:2004bf,Boucaud:2003xi}, so this behavior is not in disagreement with this possibility.

\section{Summary}\label{ssum}

Summarizing, the propagator in two and three dimensions are qualitatively, and actually also quantitatively when measured in units of the string tension, very similar, for the gauge groups SU(2), SU(3), SU(4), SU(5), SU(6) and $G_2$, and thus for the corresponding gauge algebras. In particular, no pronounced dependence on $N$ is observed for the approximately fixed value of $g^2 C_A$ used here. Hence, the dominating contribution in the mid-momentum regime and at low momenta are of order ${\cal O}(1)$ in terms of $N$ counting. In particular, for all $N$ the propagators show a behavior which is distinctively non-perturbative.

Furthermore, the results are compatible with predictions for the gauge-group dependence expected from functional calculations. In particular, the deep infrared remains the same for any gauge group.

These results emphasize that for properties of Yang-Mills theories on the level of gluonic correlation functions the specific gauge algebra is qualitatively, and to some extent even quantitatively, rather irrelevant. Only when matter is coupled to the Yang-Mills field dynamically the gauge algebra structure becomes quite relevant (see e.\ g.\ \cite{Alkofer:2010tq}). Of course, even for pure Yang-Mills theory the gauge algebra is important for some quantities, e.\ g., the order of the finite-temperature phase transition.

\acknowledgments

I am grateful to A.~Cucchieri, T.~Mendes, and {\v S}.~Olejn\'ik for helpful discussions. This work was supported by the DFG under grant  number MA 3935/1-1 and MA 3935/1-2 and by the FWF under grant number P20330 and M1099-N16. Part of the computing time was provided by the Slovak Grant Agency for Science, Grant VEGA No.\ 2/6068/2006 and by the HPC cluster in Graz. The ROOT framework \cite{Brun:1997pa} has been used in this project.

\bibliographystyle{bibstyle}
\bibliography{bib}


\end{document}